\documentstyle[11pt,aaspp4]{article}

\slugcomment{Astrophysical Journal {\bf 573}, 2002 July 1 \hspace{2.3cm} 
KUNS 1769}

\def\gtrless{\mathrel{\hbox{\rlap{\hbox
                  {\lower2pt\hbox{$<$}}}{\raise3pt\hbox{$>$}}}}}
\newcommand{\mbf}[1]{\mbox{\boldmath $#1$}}

\lefthead{Nakano et al.}
\righthead{Magnetic Flux Loss in Molecular Clouds}

\begin{document}

\title{Mechanism of Magnetic Flux Loss in Molecular Clouds}

\author{Takenori Nakano and Ryoichi Nishi\altaffilmark{1}}
\affil{Department of Physics, Graduate School of Science, Kyoto University, 
    Kyoto 606-8502, Japan}

\and

\author{Toyoharu Umebayashi}
\affil{Computing Service Center, Yamagata University, Yamagata 990-8560, 
    Japan}

\altaffiltext{1}{Address from April 2002: Department of Physics, Faculty of 
    Science, Niigata University, Niigata 950-2181, Japan}

\begin{abstract}
We investigate the detailed processes working in the drift of magnetic 
fields in molecular clouds.  To the frictional force, whereby the magnetic 
force is transmitted to neutral molecules, ions contribute more than half 
only at cloud densities $n_{\rm H} \lesssim 10^4\,{\rm cm}^{-3}$, and 
charged grains contribute more than about 90\% at 
$n_{\rm H} \gtrsim 10^6\,{\rm cm}^{-3}$.  Thus grains play a decisive role 
in the process of magnetic flux loss.  Approximating the flux loss time 
$t_B$ by a power law $t_B \propto B^{-\gamma}$, where $B$ is the mean 
field strength in the cloud, we find $\gamma \approx 2$, characteristic 
to ambipolar diffusion, only at $n_{\rm H} \lesssim 10^7\,{\rm cm}^{-3}$ 
where ions and smallest grains are pretty well frozen to magnetic fields.  
At $n_{\rm H} > 10^7\,{\rm cm}^{-3}$, $\gamma$ decreases steeply with 
$n_{\rm H}$, and finally at 
    $n_{\rm H} \approx n_{\rm dec} \approx
                  {\rm a \,\,few} \times 10^{11}\,{\rm cm}^{-3}$, 
where magnetic fields effectively decouple from the gas, $\gamma \ll 1$ 
is attained, reminiscent of Ohmic dissipation, though flux loss occurs 
about 10 times faster than by pure Ohmic dissipation.  Because even ions 
are not very well frozen at $n_{\rm H} > 10^7\,{\rm cm}^{-3}$,  
ions and grains drift slower than magnetic fields.  This insufficient 
freezing makes $t_B$ more and more insensitive to $B$ as $n_{\rm H}$ 
increases.  Ohmic dissipation is dominant only at 
    $n_{\rm H} \gtrsim 1 \times 10^{12}\,{\rm cm}^{-3}$.  
While ions and electrons drift in the direction of magnetic force at all 
densities, grains of opposite charges drift in opposite directions at high 
densities, where grains are major contributors to the frictional force.  
Although magnetic flux loss occurs significantly faster than by Ohmic 
dissipation even at very high densities as $n_{\rm H} \approx n_{\rm dec}$, 
the process going on at high densities is quite different from ambipolar 
diffusion in which particles of opposite charges are supposed to drift 
as one unit.
\end{abstract}

\keywords{ISM: clouds --- ISM: dust --- ISM: magnetic fields --- magnetic 
fields --- plasmas --- stars: formation}

\section{Introduction}

Magnetic fields of interstellar molecular clouds are widely believed to have 
significant effect on star formation.  For an oblate cloud or cloud core of 
mass $M$ contracted along field lines to some extent, there is a critical 
value for its magnetic flux ${\mit\Phi}$ given by
\begin{equation}
   {\mit\Phi}_{\rm cr} = f_\phi G^{1/2} M,  \label{phicr}
\end{equation}
where $G$ is the gravitational constant and $f_\phi$ is a dimensionless 
constant.  A cloud (core) with  ${\mit\Phi} < {\mit\Phi}_{\rm cr}$  cannot 
be kept in hydrostatic equilibrium by the magnetic force alone (this state 
is widely called magnetically supercritical), and one with  
${\mit\Phi} > {\mit\Phi}_{\rm cr}$ (magnetically subcritical) can be in 
equilibrium if its expansion is suppressed by external magnetic fields.  
Applying the virial theorem to such clouds, Strittmatter (1966) found 
$f_\phi$ between 4.9 and 9.4 depending on the flatness of the cloud.  
From some numerical cloud models Mouschovias \& Spitzer (1976) obtained  
$f_\phi \approx 8.0$,  and Tomisaka, Ikeuchi, \& Nakamura (1988) got  
$f_\phi \approx 8.3$.  Li \& Shu (1996) found $f_\phi = 2 \pi$ for
self-similar, singular, isothermal clouds.

We can define the critical magnetic field strength of the cloud (core) by
\begin{equation}
   B_{\rm cr} = \frac{{\mit\Phi}_{\rm cr}}{\pi R^2}
              = f_\phi G^{1/2} {\mit\Sigma},   \label{bcr}
\end{equation}
where $R$ is the radius of the cloud (core) perpendicular to the mean 
magnetic field direction and ${\mit\Sigma} = M / \pi R^2$ is the mean 
column density of the cloud (core) along field lines.  Of course 
$B \gtrless B_{\rm cr}$ is equivalent to  
${\mit\Phi} \gtrless {\mit\Phi}_{\rm cr}$, where $B = {\mit\Phi}/ \pi R^2$ 
is the mean field strength in the cloud (core).  Nakano \& Nakamura (1978) 
found that isothermal disks penetrated by uniform magnetic fields 
$\mbf{B}$ perpendicular to the disk layers are gravitationally unstable 
only when $B < B_{\rm cr}$, where $B_{\rm cr}$ is given by equation 
(\ref{bcr}) with $f_\phi = 2 \pi$.  The critical wavelength, below which 
the disk is unstable, decreases as $B/B_{\rm cr}$ decreases (\cite{nak88}).  
These mean that with perturbations at least the disk as a whole can 
contract perpendicular to field lines when $B < B_{\rm cr}$.

While magnetic fluxes of clouds estimated by observations are not much 
smaller than their critical values, or 
${\mit\Phi}/{\mit\Phi}_{\rm cr} \sim 1$  (e.g., \cite{cru99}), magnetic 
stars with mean surface fields of 1\,kG to 30\,kG have ratios  
${\mit\Phi} / {\mit\Phi}_{\rm cr} \approx 10^{-5} - 10^{-3}$  
(\cite{nak83}).  For ordinary stars like the Sun with mean surface fields 
of  $\sim 1$\,G, ${\mit\Phi} / {\mit\Phi}_{\rm cr}$  is as 
small as $10^{-8}$.  This suggests that cloud cores must lose most of 
their initial magnetic fluxes at some stages of star formation.  At what 
stages and by what mechanisms\,?  This is called the magnetic flux problem 
in star formation (e.g., \cite{nak84}).

Ohmic dissipation is too slow to dissipate magnetic fields in molecular 
clouds of ordinary densities.  For example, for the ionization fraction 
$10^{-8}$ at density $10^5\,{\rm cm}^{-3}$ (Figure 1) and the length scale 
of magnetic fields, $0.1$\,pc, about the Jeans length at this density and 
temperature $10\,$K, we obtain the Ohmic dissipation time $10^{15}$\,yr, 
larger than the age of the universe by orders of magnitude.  
Mestel \& Spitzer (1956) found another process of decreasing magnetic 
flux, which is now widely called ambipolar diffusion (sometimes called 
plasma drift; see Appendix~\ref{termin} for the terminology) and is much 
more efficient than Ohmic dissipation in molecular clouds of ordinary 
densities.  In this process ions (and electrons), which are well frozen 
to magnetic fields, drift in the sea of neutral molecules together with 
magnetic fields at a terminal velocity with which the magnetic force 
balances with the frictional force exerted by the neutrals.

Because some dust grains in clouds are electrically charged and interact 
with magnetic fields, they contribute to controlling the drift of magnetic 
fields as well as ions.  However, because grains are not so strongly 
coupled with magnetic fields as ions due to their large masses, complete 
freezing is not a good approximation in most situations.  More accurate 
treatment is required for their motion (\cite{elm79}).  Moreover, the size 
and the mass of grains are distributed in wide ranges (e.g., Mathis, Rumpl, 
\& Nordsieck 1977, referred to as MRN hereafter), and the degree of 
freezing depends sensitively on their masses.  More accurate treatment is 
also necessary for ions because even ions are not well frozen at high 
densities.  Nakano (1984) and Nakano \& Umebayashi (1986, referred to as 
\cite{nak86} hereafter) formulated a method of describing the drift of 
magnetic fields in clouds containing any kinds of charged particles 
which are coupled with magnetic fields at arbitrary strengths.

Using this formalism we investigated the time scale of magnetic flux loss, 
$t_B$, from a major part of a cloud (core) as a function of the mean 
density of the cloud (core) (\cite{nak84}; \cite{nak86}; \cite{ume90}; 
Nishi, Nakano, \& Umebayashi 1991, referred to as \cite{nis91} hereafter).  
We have found that there is a critical value $n_{\rm dec} \approx 10^{11}$ 
hydrogen nuclei per ${\rm cm}^3$ for the density of the cloud, $n_{\rm H}$, 
at which $t_B$ is equal to the free-fall time $t_{\rm ff}$ of the cloud 
(core);  $t_B < t_{\rm ff}$ holds only at $n_{\rm H} > n_{\rm dec}$, 
and $t_B \gg t_{\rm ff}$ at $n_{\rm H} \ll n_{\rm dec}$.  We have called 
this the decoupling density because extensive flux loss occurs only at 
$n_{\rm H} \gtrsim n_{\rm dec}$.

Our previous results show that at 
$10^3\,{\rm cm}^{-3} \lesssim n_{\rm H} \lesssim 10^7\,{\rm cm}^{-3}$, 
$t_B$ is 10 to $10^2$ times $t_{\rm ff}$ for $B = B_{\rm cr}$ and 
a relation $t_B \propto B^{-2}$ approximately holds at least for 
$B_{\rm cr} \geq B \gtrsim 0.1 B_{\rm cr}$.  The relation 
$t_B \propto B^{-2}$ is characteristic to ambipolar diffusion which 
occurs when the dominant charged particles are well frozen to magnetic 
fields (see \S\,\ref{limitingcase} for details).  As the density increases 
at $n_{\rm H} \gtrsim 10^7\,{\rm cm}^{-3}$, $t_B$ becomes more and more 
insensitive to $B$, and finally at $n_{\rm H} \approx n_{\rm dec}$, 
$t_B$ becomes almost independent of $B$, reminiscent of Ohmic dissipation.  
Recently Desch \& Mouschovias (2001) wrote that ambipolar diffusion was 
the dominant process even at densities several orders of magnitude higher 
than $n_{\rm dec}$.  These results may cause confusion.  Besides, the 
dependence of $t_B$ on $n_{\rm H}$ and $B$ described above is rather 
complicated.  It would be necessary to clarify what is going on especially 
at high densities.  Furthermore, although it was pointed out that grains 
play an important role in the process of magnetic flux loss, it does not 
seem to be widely recognized how important they are.

The purpose of this paper is to clarify detailed mechanisms operating in 
the loss of magnetic flux in molecular clouds especially at high densities  
and to show how the grains behave.  In \S\,2 we summarize some of the 
formulae obtained by Nakano (1984) and \cite{nak86}, which will be used 
in this paper.  In \S\,3 we show numerical results and analyze in detail 
the processes going on especially at high densities.  Discussion is made 
in \S\,4, and summary is given in \S\,5.  In Appendix~\ref{anotherform} 
we show another method of the formulation than that of Nakano (1984) and 
\cite{nak86}.

\section{Drift of Charged Particles and Magnetic Fields} \label{drift}

\subsection{Formulae} \label{formulae}

We summarize some of the formulae obtained by Nakano (1984) and 
\cite{nak86}, which will be used in this paper.  We consider a cloud 
which is composed mainly of neutral molecules and atoms but contains 
a slight amount of charged particles such as various atomic and molecular 
ions, electrons, and grains of various sizes and charges.

Because each kind of charged particles are scarce, we can neglect in 
its equation of motion the pressure force, the gravity, and the inertia 
term compared with the Lorentz force and the frictional force exerted by 
the neutrals.  With this approximation we obtain the drift velocity 
(velocity of the guiding center)  $\mbf{v}_\lambda$  of an arbitrary 
particle $\lambda$ of mass $m_\lambda$ and electric charge $q_\lambda$ 
relative to the neutrals.  We adopt the local Cartesian coordinate system 
with the $z$-axis parallel to the local magnetic field vector $\mbf{B}$ 
and the $x$-axis along the magnetic force $\mbf{j \times B}/c$, where 
$\mbf{j} = (c/4\pi)\mbf{\nabla \times B}$ is the electric current density 
and $c$ is the light velocity.  The components of the drift velocity 
$\mbf{v}_\lambda$ are given by
\begin{equation}
   v_{\lambda x} = \frac{(\tau_\lambda \omega_\lambda)^2}
                        {1 + (\tau_\lambda \omega_\lambda)^2} \,
                   \frac{1}{A_1^2 + A_2^2} \,
               \bigg(A_1 + \frac{A_2}{\tau_\lambda \omega_\lambda}\bigg) \,
               \frac{1}{c} \,|\,\mbf{j \times B}|, \label{vlx}
\end{equation}
\begin{equation}
   v_{\lambda y} = \frac{(\tau_\lambda \omega_\lambda)^2}
                        {1 + (\tau_\lambda \omega_\lambda)^2} \,
                   \frac{1}{A_1^2 + A_2^2} \,
               \bigg(\frac{A_1}{\tau_\lambda \omega_\lambda} - A_2\bigg) \,
               \frac{1}{c} \,|\,\mbf{j \times B}|. \label{vly}
\end{equation}
Here, $\tau_\lambda$ is the viscous damping time of the motion of particle 
$\lambda$ in the sea of the neutrals, whose expressions are given, e.g., 
by Nakano (1984) for various particles,
\begin{equation}
   \omega_\lambda = \frac{q_\lambda B}{m_\lambda c}    \label{omegl}
\end{equation}
is the gyrofrequency (defined to be negative for negatively charged 
particles),
\begin{equation}
   A_1 = \sum_\nu \frac{\rho_\nu \tau_\nu \omega_\nu^2}
                       {1 + (\tau_\nu \omega_\nu)^2}
       = \frac{B}{c} \sum_\nu \frac{n_\nu q_\nu \tau_\nu \omega_\nu}
                       {1 + (\tau_\nu \omega_\nu)^2}
       = \bigg(\frac{B}{c}\bigg)^2 \sigma_{\rm P},      \label{a1}
\end{equation}
\begin{equation}
   A_2 = \sum_\nu \frac{\rho_\nu \omega_\nu}
                       {1 + (\tau_\nu \omega_\nu)^2}
       = \frac{B}{c} \sum_\nu \frac{n_\nu q_\nu}
                       {1 + (\tau_\nu \omega_\nu)^2}
       = \bigg(\frac{B}{c}\bigg)^2 \sigma_{\rm H},      \label{a2}
\end{equation}
where $\rho_\nu$ and $n_\nu = \rho_\nu/m_\nu$ are the mass and number 
densities, respectively, of particle $\nu$, and $\sigma_{\rm P}$ and 
$\sigma_{\rm H}$ are Pedersen and Hall conductivities, respectively (see 
Appendix~\ref{anotherform}).  Summation in equations (\ref{a1}) and 
(\ref{a2}) is for all kinds of charged particles; particles having different 
values of $\tau_\nu$ or $\omega_\nu$ are different kinds.  The component 
$v_{\lambda z}$ is not necessary because the drift along field lines has 
no effect on the magnetic flux loss.

The drift velocity of magnetic fields, $\mbf{v}_B$, is defined as the 
velocity relative to the neutrals of each point on an arbitrary closed 
contour in the cloud with which the magnetic flux through the contour is 
conserved.  This requires
\begin{equation}
   \mbf{E}_\perp + \frac{1}{c}\, \mbf{v}_B \mbf{\times B} = 0,
                                                \label{elecfield} 
\end{equation}
where $\mbf{E}_\perp$ is the component of the electric field $\mbf{E}$ 
perpendicular to $\mbf{B}$ in the frame moving with the neutrals.  Using 
the relation between $\mbf{v}_\lambda$ and $\mbf{E}$, or the equation of 
motion, we obtain
\begin{equation}
   v_{Bx} =  \frac{A_1}{A_1^2 + A_2^2} \,
             \frac{1}{c} \, |\,\mbf{j \times B}|,   \label{vbx}
\end{equation}
\begin{equation}
   v_{By} = -\frac{A_2}{A_1^2 + A_2^2} \,
             \frac{1}{c} \, |\,\mbf{j \times B}|.   \label{vby}
\end{equation}
The rate of magnetic flux loss is determined by $v_{Bx}$ alone 
(\cite{nak86}).  This is self-evident for axisymmetric clouds because the 
local $y$-axis is in the azimuthal direction.  Therefore, the subscript 
$x$ of $v_{Bx}$ and $v_{\lambda x}$ may be omitted in the following.

Using equations (\ref{vlx}) and (\ref{vly}) we can calculate the electric 
current density in the ($x, y$) plane, 
     $\mbf{J} = \sum_\lambda n_\lambda q_\lambda \mbf{v}_\lambda$.  
With the electrical neutrality relation 
     $\sum_\lambda \rho_\lambda \omega_\lambda = 0$, 
we can easily show that $J_x = 0$, which is consistent with the definition 
of the $x$-axis, and that $J_y$ is equal to the component perpendicular to 
$\mbf{B}$ of $\mbf{j}$ which appears in equations (\ref{vlx}) and 
(\ref{vly}).  Thus the formulation has been consistently done.  Consistency 
of equations (\ref{vbx}) and (\ref{vby}) with equations (\ref{vlx}) and 
(\ref{vly}) can be confirmed by considering the motion of a test particle 
which is completely frozen to magnetic fields and drifts with magnetic 
fields.  Taking a limit 
     $|\tau_\lambda \omega_\lambda| \rightarrow \infty$ 
reduces equations (\ref{vlx}) and (\ref{vly}) to equations (\ref{vbx}) and 
(\ref{vby}), respectively, because $A_1$ and $A_2$ are not affected by the 
test particle.  

In this formalism we have neglected the effect of charge fluctuation of 
grains caused by sticking of ions and electrons on their motion because 
Nakano \& Umebayashi (1980) and \cite{nak86} had a rough estimation that 
this effect is small.  More elaborate discussion on this effect will be 
given in \S\,\ref{cfluct}.  We have also neglected the collision between 
charged particles, whose effect is negligibly small compared with the 
effect of their collision with the neutrals as shown in 
Appendix~\ref{collision}.

With the approximation made on the motion of charged particles, the 
magnetic force on charged particles must balance with the frictional 
force exerted on them by the neutrals, or by components
\begin{equation}
   \sum_\lambda \frac{\rho_\lambda v_{\lambda x}}{\tau_\lambda}
      = \frac{1}{c} \, |\,\mbf{j \times B}|,   \label{frictionx}
\end{equation}
\begin{equation}
   \sum_\lambda \frac{\rho_\lambda v_{\lambda y}}{\tau_\lambda} = 0.
                                                \label{frictiony}
\end{equation}
It is easy to confirm that $\mbf{v}_\lambda$ given by equations (\ref{vlx}) 
and (\ref{vly}) satisfies these equations.

One may simply think that the quantity $| \tau_\lambda \omega_\lambda |$ 
would characterize the degree of freezing of particle $\lambda$ to magnetic 
fields.  However, we have from equations (\ref{vlx}) and (\ref{vbx})
\begin{equation}
   \frac{v_{\lambda x}}{v_{Bx}} = \frac{(\tau_\lambda \omega_\lambda)^2}
                                   {1 + (\tau_\lambda \omega_\lambda)^2}
             \bigg(1 + \frac{A_2}{A_1 \tau_\lambda \omega_\lambda}\bigg).
                                                         \label{vlvb}
\end{equation}
When $| \tau_\lambda \omega_\lambda | \gg 1$, the degree of freezing, 
or the relative drift velocity, is certainly determined by 
$\tau_\lambda \omega_\lambda$ alone as 
$v_{\lambda x}/v_{Bx} \approx (\tau_\lambda \omega_\lambda)^2 / 
     [1 + (\tau_\lambda \omega_\lambda)^2]$ 
because $|A_2| / A_1$ is at most several at least in the ranges of density 
and field strength covered by this paper.  However, when  
$|\tau_\lambda \omega_\lambda|$ is not much larger than 1, the relative 
drift velocity may be greatly affected by the second term in the 
parentheses of equation (\ref{vlvb}), or by other charged particles.  
We shall show some examples of this effect in \S\S\,\ref{bdependence} and 
\ref{contgrain}.

Making use of equations (\ref{elecfield}), (\ref{vbx}), and (\ref{vby}), 
\cite{nak86} showed that the electric field 
     $\mbf{E}_0 = \mbf{E - u_{\rm n} \times B}/c$  
in the frame wherein the neutrals move with a velocity  $\mbf{u}_{\rm n}$ 
is given by
\begin{equation}
   \mbf{E}_0 = - \frac{1}{c} \, \mbf{u_{\rm n} \times B}
               + \frac{1}{\sigma_\parallel} \, \mbf{j}
               + \beta \, \mbf{j \times B}
               - \xi \, \mbf{(j \times B) \times B}, \label{e0}
\end{equation}
where
\begin{equation}
   \sigma_\parallel = \sum_\nu \sigma_\nu, \quad
   \sigma_\nu = \frac{n_\nu q_\nu^2 \tau_\nu}{m_\nu}
         = \frac{c}{B} n_\nu q_\nu \tau_\nu \omega_\nu   \label{sigmapara}
\end{equation}
is the electric conductivity parallel to magnetic field lines, and
\begin{equation}
   \beta = \frac{B}{c^2} \, \frac{A_2}{A_1^2 + A_2^2},   \label{beta}
\end{equation}
\begin{equation}
   \xi = \frac{1}{c^2} \, \frac{A_1}{A_1^2 + A_2^2}
         - \frac{1}{B^2 \sigma_\parallel}.   \label{xi}
\end{equation}
The work $\mbf{j \cdot E}_0$ gives the dissipation rate of magnetic energy 
per unit volume.  The first term on the right-hand side of equation 
(\ref{e0}) causes amplification of magnetic fields by fluid motion.  
The second term leads to Ohmic dissipation.  The third term yields no 
work.  The last term gives rise to the dissipation in excess of Ohmic 
dissipation;  ambipolar diffusion in excess of Ohmic dissipation is given 
by this term when the dominant charged particles are well frozen to 
magnetic fields.  Thus the ratio of the last term to the second term in 
equation (\ref{e0})
\begin{equation}
   D \equiv B^2 \xi \sigma_\parallel
     = \frac{\sigma_\parallel \sigma_{\rm P}}
            {\sigma_{\rm P}^2 + \sigma_{\rm H}^2} - 1   \label{excess}
\end{equation}
gives the excess dissipation relative to Ohmic dissipation (\cite{nak86}).

In Appendix~\ref{anotherform} we show another method of obtaining these 
formulae than that adopted by Nakano (1984) and \cite{nak86}.

\subsection{Limiting Cases} \label{limitingcase}

Because the drift velocity $v_{Bx}$ is expressed in terms of the 
complicated quantities $A_1$ and $A_2$, it would be worthwhile to show how 
magnetic fields drift in some limiting cases.

When the dominant charged particles are well frozen to magnetic fields, or 
$| \tau_\nu \omega_\nu | \gg 1$, equations (\ref{a1}) and (\ref{a2}) give 
    $A_1 \approx  \sum_\nu \rho_\nu / \tau_\nu    \gg 
   |A_2| \approx |\sum_\nu \rho_\nu /(\tau_\nu^2 \omega_\nu)|$, 
and thus these particles drift at almost the same velocity irrespective of 
their charges as seen from equation (\ref{vlx}).  These relations also 
reduce equation (\ref{vbx}) to
\begin{equation}
   v_{Bx} \sum_\nu \frac{\rho_\nu}{\tau_\nu} 
        \approx \frac{1}{c} \, |\,\mbf{j \times B}|.  \label{vbambi}
\end{equation}
This can also be obtained by replacing $v_{\lambda x}$ with $v_{Bx}$ in 
equation (\ref{frictionx}).  Equation (\ref{vbambi}) means that the charged 
particles drift together with magnetic fields with the terminal velocity 
at which the magnetic force balances with the frictional force exerted by 
the neutrals.  This is the generalization of ambipolar diffusion 
investigated first by Mestel \& Spitzer (1956), who considered only ions 
as the transmitter of the magnetic force to the neutrals.  Because the 
magnetic force balances with the gravitational force perpendicular to field 
lines when $B = B_{\rm cr}$, the mean magnetic force in an oblate cloud 
with $B \leq B_{\rm cr}$ can be given by
\begin{equation}
   \frac{1}{c} \, |\,\mbf{j \times B}| \approx 
        \bigg(\frac{B}{B_{\rm cr}}\bigg)^2 \,\frac{\pi}{2}
                           G {\mit\Sigma} \rho,        \label{magforce}
\end{equation}
where $\rho$ and ${\mit\Sigma}$ are the mean density and the mean column 
density along field lines, respectively, of the cloud.  From equations 
(\ref{vbambi}) and (\ref{magforce}) we have $v_{Bx} \propto B^2$, and then 
the magnetic flux loss time from the major part of the cloud, defined as 
the time required to drift a length scale $L$ of magnetic fields, 
$t_B \approx L/v_{Bx} \propto B^{-2}$.  This is characteristic to ambipolar 
diffusion in magnetically supercritical clouds.

When ions are the major contributors to the left-hand side of equation 
(\ref{vbambi}), whose mass $m_{\rm i}$ is much larger than that of 
an H$_2$ molecule and whose viscous damping time is given by 
$\tau_{\rm i} \approx m_{\rm i} / (\rho \langle\sigma v \rangle_{\rm i})$ 
with the collision rate coefficient 
     $\langle\sigma v \rangle_{\rm i} \approx 
               1.5 \times 10^{-9}\,{\rm cm}^3 \,{\rm s}^{-1}$ 
(\cite{nak84}), we have
\begin{equation}
  t_B \approx \frac{n_{\rm i} \langle\sigma v \rangle_{\rm i}}{\pi G \rho}
               \bigg( \frac{B_{\rm cr}}{B} \bigg)^2   \label{tbambi}
\end{equation}
for nearly spherical clouds with $\rho \approx {\mit\Sigma}/(2L)$ insofar 
as $B \leq B_{\rm cr}$.  This is a well-known expression of the ambipolar 
diffusion time for magnetically supercritical clouds, which is proportional 
to the ionization fraction  $n_{\rm i}/n_{\rm H}$.

If there exists a magnetically subcritical cloud core ($B > B_{\rm cr}$) 
in equilibrium owing to external magnetic fields which suppress its 
expansion, the magnetic force therein almost balances with the gravity, or 
    $|\,\protect\mbf{j \times B}|/c \approx \pi G {\mit\Sigma} \rho /2$ 
though this must be significantly smaller than $B^2/(4\pi R)$, where $R$ 
is the core radius.  Substituting this relation into equation 
(\ref{vbambi}), we find that $v_{Bx}$ and $t_B$ are independent of $B$ 
and equal to those at $B = B_{\rm cr}$ as shown by Nakano (1998).

Another limiting case is where the dominant charged particles are not 
at all frozen to magnetic fields, or $| \tau_\nu \omega_\nu | \ll 1$.  
In this case we have
\begin{equation}
   A_1 \approx \sum_\nu \rho_\nu \tau_\nu \omega_\nu^2
           = \bigg( \frac{B}{c} \bigg)^2 \sigma_\parallel,  \label{a1nf}
\end{equation}
\begin{equation}
   A_2 \approx \sum_\nu \rho_\nu \omega_\nu (1 - \tau_\nu^2 \omega_\nu^2)
            = -\sum_\nu \rho_\nu \tau_\nu^2\omega_\nu^3.  \label{a2nf}
\end{equation}
Electrical neutrality derives the last expression of equation (\ref{a2nf}).  
Again we find $A_1 \gg | A_2 |$.  Taking 
$|\, \mbf{j \times B}| /c \approx B^2 /(4\pi L)$, we obtain the drift time 
of magnetic fields
\begin{equation}
   t_B \approx \frac{L}{v_{Bx}}
       \approx \frac{4\pi \sigma_\parallel}{c^2} L^2.  \label{tbod}
\end{equation}
This is the well-known Ohmic dissipation time, which does not depend on 
the field strength.

Thus our formalism contains ambipolar diffusion and Ohmic dissipation as 
limiting cases.

\section{Numerical Results}

\subsection{Cloud Model}

We adopt the cloud model almost the same as \cite{nis91}.  Because 
magnetized clouds can contract along field lines rather easily, the force 
balance may approximately hold between gravity and gas pressure along 
field lines even when they are contracting dynamically perpendicular to 
field lines (e.g., \cite{sco80}; Nakamura, Hanawa, \& Nakano 1995).  
In such clouds the mean density $\rho$ and the mean column density 
${\mit\Sigma}$ along field lines have a relation
\begin{equation}
   {\mit\Sigma} \approx \bigg(\frac{4kT\rho}{\pi G\mu m_{\rm H}}\bigg)^{1/2}
        \approx 0.040 \bigg(\frac{n_{\rm H}}{10^5\,{\rm cm}^{-3}}\,
                        \frac{T}{10\,{\rm K}}\,\frac{2.37}{\mu}\bigg)^{1/2}
                \,{\rm g \,cm}^{-2},     \label{sigma}
\end{equation}
where $T$ is the mean temperature of the cloud, $k$ is the Boltzmann 
constant, $\mu$ is the mean molecular weight of the gas, and $m_{\rm H}$ 
is the mass of a hydrogen atom.  Equation (\ref{sigma}) gives the column 
density of an isothermal disk in equilibrium if $\rho$ is half the 
density at the midplane.  If magnetic fields are weak and the isothermal 
(spherical) cloud is on the verge of collapse (\cite{ebe55}; \cite{bon56}), 
the mean column density is $0.9$ times that of equation (\ref{sigma}).  
The half-thickness of the cloud is given by 
$Z \approx {\mit\Sigma} /(2\rho)$, or
\begin{equation}
   Z \approx \bigg( \frac{kT}{\pi G \mu m_{\rm H} \rho} \bigg)^{1/2}
     \approx 0.027 \bigg(\frac{10^5\,{\rm cm}^{-3}}{n_{\rm H}}\,
                       \frac{T}{10\,{\rm K}}\,\frac{2.37}{\mu} \bigg)^{1/2}
                               \,{\rm pc}.        \label{thickness}
\end{equation}
Both ${\mit\Sigma}$ and $Z$ are independent of the cloud mass.

\subsection{Densities of Charged Particles}

As the reaction scheme of determining densities of various charged 
particles, we adopt the same model as Umebayashi \& Nakano (1990) and 
\cite{nis91} though we have revised some of the rate coefficients 
according to Le Teuff, Millar, \& Markwick (2000);  the greatest change is 
for dissociative recombination of H$_3^+$.  In dense clouds 
sufficiently opaque to interstellar ultraviolet radiation, ions and 
free electrons are formed mainly by ionization of H$_2$ molecules and He 
atoms by cosmic rays.  After some reactions in the gas phase and at grain 
surface they finally recombine each other.  We determine the densities of 
various charged particles assuming steady state for all the reactions.

As cosmic rays go deep into a cloud, their intensity decreases by 
interaction with matter, and the ionization rate decreases 
exponentially as $\zeta = \zeta_0 \exp(-\chi/\chi_0)$, where $\zeta_0$ is 
the ionization rate of an H$_2$ molecule at the cloud surface, $\chi$ is 
the depth in column density from the cloud surface, and 
$\chi_0 \approx 96\,{\rm g\,cm}^{-2}$ is the attenuation length of the 
ionization rate (\cite{ume81}).  We take $\chi = {\mit\Sigma}/4$ because 
we are interested in the mean densities of various particles in the cloud.  
We also take into account the ionization by radioactive elements contained 
in the cloud at a rate $6.9 \times 10^{-23}\,{\rm s}^{-1}$ (\cite{ume81}), 
which is important only at $n_{\rm H} \gtrsim 10^{15}\,{\rm cm}^{-3}$.

As charged particles we consider electrons e$^-$, atomic ions H$^+$, 
He$^+$, and C$^+$, metallic ions such as Mg$^+$, Si$^+$, and Fe$^+$, 
which we denote as M$^+$ collectively, H$_3^+$, molecular ions other than 
H$_3^+$ (typically HCO$^+$), which we denote as m$^+$, and charged grains.  
As the mean masses of M$^+$ and m$^+$ we take $34m_{\rm H}$ and 
$29m_{\rm H}$, respectively.  We separate H$_3^+$ from m$^+$ because the 
large difference in their masses causes considerable differences in 
$\tau_\lambda$ and $\omega_\lambda$.  We adopt the MRN size distribution 
of grains given by
\begin{equation}
   \frac{dn_{\rm g}}{da} = C n_{\rm H} a^{-3.5}, \quad
                a_{\rm min} \leq a \leq a_{\rm max},  \label{dngdna}
\end{equation}
where $a$ is the grain radius and 
$C \approx 1.5 \times 10^{-25}\,{\rm cm}^{2.5}$ (\cite{dra84}; 
\cite{mat86}).  We take $a_{\rm min} = 5$\,nm and 
$a_{\rm max} = 250$\,nm.  As for the charge states of grains we consider 
neutral, $\pm e$, $\pm 2e$, and $\pm 3e$, where $e$ is the elementary 
electric charge.  We divide the grain radius into 40 bins of equal 
logarithmic width ${\mit\Delta} \log a = 0.0425$.  This must be accurate 
enough for our purpose because we have found little difference in 
numerical results compared with the case of 20 bins adopted by 
\cite{nis91}\footnote{
   Desch \& Mouschovias (2001) criticized \cite{nis91} in their \S\,2.1 
   claiming (i) ignorance of the size-dependence of grain charges, and 
   (ii) ignorance of collisions between grains of different sizes which 
   is important to neutralization of grain charges at high densities.  
   In reality, however, \cite{nis91} did not make these ignorances.  As for 
   (i) see their Appendix~A, where the equations on grain charges have 
   clear dependences on the grain radius.  \cite{nis91} did not write 
   anything that means (ii).},  
though we have found some difference from the case of 5 bins adopted by 
Desch \& Mouschovias (2001).

Figure 1 shows abundances of various particles as functions of the 
cloud density $n_{\rm H}$ for the {\it standard} case 
$\zeta_0 = 1 \times 10^{-17}\,{\rm s}^{-1}$ (\cite{spi68}).  Ions and 
electrons are dominant charged particles at 
$n_{\rm H} \lesssim 10^6\,{\rm cm}^{-3}$, ions and g$^-$ (grains of charge 
$-e$) are dominant at $n_{\rm H}$ between $10^7$ and $10^9\,{\rm cm}^{-3}$, 
and g$^-$ and g$^+$ are major at 
$n_{\rm H} \gtrsim 10^{10}\,{\rm cm}^{-3}$.  To keep electrical neutrality 
at very high densities $n_{\rm H} \gtrsim 10^{10}\,{\rm cm}^{-3}$  where 
ions and electrons are no longer dominant charged particles because of 
their efficient recombination, $n({\rm g}^-) \approx n({\rm g}^+)$ must 
hold.  Nakano (1984) showed that to assure this relation the number flux 
of thermal electrons must be $(S_{\rm i}/S_{\rm e})^{1/2}$ times that of 
thermal ions, or
\begin{equation}
   \frac{n_{\rm e}}{n_{\rm i}} \approx \bigg(\frac{S_{\rm i}}{S_{\rm e}}
                \frac{m_{\rm e}}{m_{\rm i}} \bigg)^{1/2},   \label{neni}
\end{equation}
where $S_{\rm i}$ and $S_{\rm e}$ are sticking probabilities of ions and 
electrons, respectively, when they collide neutral grains.  Although 
Nakano (1984) ignored electric polarization of grains induced by 
approaching charged particles, which has an effect of enhancing the 
collision rate (\cite{dra87}), this effect cancels out in the derivation 
of equation (\ref{neni}).  As seen in Figure 1, metallic ions are 
dominant among various ions, or $n_{\rm i} \approx n({\rm M}^+)$, at all 
densities.  In numerical calculation we took $S_{\rm e} = 0.6$ and 
$S_{\rm i} = 1$ (\cite{ume80}; \cite{nis91}).  This is why 
$n_{\rm e} \approx 6 \times 10^{-3} n({\rm M}^+)$ holds at 
$n_{\rm H} \gtrsim 10^{10}\,{\rm cm}^{-3}$.  Equation (\ref{neni}) 
also yields $n({\rm g}^{--}) \approx n({\rm g}^{++})$.  Decrease of 
charged grains with $n_{\rm H}$ at very high densities is due to 
neutralization of grains by mutual collisions.  Cosmic rays are 
significantly attenuated at $n_{\rm H} \gtrsim 10^{12}\,{\rm cm}^{-3}$.

Figure 2 shows the charge state of grains as a function of the grain 
radius at several cloud densities.  At low densities 
$n_{\rm H} \lesssim 10^6\,{\rm cm}^{-3}$ grains of charge $-e$ are dominant 
and $1/4$ to $1/30$ of grains are neutral depending on their radius.  
At $n_{\rm H} \approx 10^8\,{\rm cm}^{-3}$ neutral grains are as abundant 
as $-e$ grains and grains of other charges are much less at all radii.  
At very high densities $n_{\rm H} \gtrsim 10^{10}\,{\rm cm}^{-3}$ neutral 
grains are most abundant, and $+e$ grains are as abundant as $-e$ grains 
at all radii.

\subsection{Magnetic Flux Loss Time} \label{losstime}

The mean magnetic force in an oblate cloud with $B \leq B_{\rm cr}$ 
satisfies equation (\ref{magforce}) whether or not the force balance holds 
along field lines.  The magnetic flux loss time $t_B$ can be given 
by the time for field lines to drift the length scale $L$ of the cloud.  
From equations (\ref{vbx}) and (\ref{magforce}) we have
\begin{equation}
   t_B \approx \frac{L}{v_{Bx}}
       \approx \bigg( \frac{B_{\rm cr}}{B} \bigg)^2
             \frac{A_1^2 + A_2^2}{A_1} \frac{2L}{\pi G {\mit\Sigma} \rho}.
                                                            \label{tb}
\end{equation}

First we consider a cloud wherein the force balance approximately holds 
between gravity and pressure along field lines and whose column density 
${\mit\Sigma}$ and half-thickness $Z$ are given by equations (\ref{sigma}) 
and (\ref{thickness}), respectively.  From equations (\ref{bcr}) and 
(\ref{sigma}) we have
\begin{equation}
   B_{\rm cr} \approx 4 \bigg(\frac{\pi kT \rho}
                               {\mu m_{\rm H}}\bigg)^{1/2}
        \approx \,6.4 \times 10^{-5}\bigg(\frac{T}{10\,{\rm K}}\,
                   \frac{n_{\rm H}}{10^5\,{\rm cm}^{-3}}\,
                   \frac{2.37}{\mu}  \bigg)^{1/2}\,{\rm G}.  \label{bcra}
\end{equation}
The quasistatic contraction of clouds induced by the drift of magnetic 
fields is highly nonhomologous;  only the densest central part of 
the cloud contracts leaving the outer part almost unchanged (\cite{nak79}, 
1982; \cite{liz89}).  The time scale of such contraction is the drift time 
of magnetic fields in the central part of the cloud whose length scale $L$ 
is nearly equal to the thickness of the cloud, $Z$, or
\begin{equation}
   t_B \approx \bigg( \frac{B_{\rm cr}}{B} \bigg)^2
               \frac{A_1^2 + A_2^2}{A_1} \frac{1}{\pi G \rho^2}.
                                                       \label{tbequil}
\end{equation}
Disk-like clouds in runaway collapse with field lines perpendicular to 
the disk layers have nearly uniform cores whose column density along field 
lines and size across them are given approximately by equations 
(\ref{sigma}) and (\ref{thickness}), respectively (\cite{nak95}).  The flux 
loss time of these cores is also given by equation (\ref{tbequil}).  This 
equation also holds for nearly spherical clouds wherein the pressure force 
almost balances with the gravity (though $B^2$ must be significantly 
smaller than $B_{\rm cr}^2$) because the cloud radius is nearly equal to 
the half-thickness along field lines, $Z$.

As mentioned above, equation (\ref{tb}) for $t_B$ holds for dynamically 
contracting clouds even if the force balance does not hold along field 
lines.  How about equation (\ref{tbequil})?  If the cloud is nearly 
spherical, equation (\ref{tbequil}) can also be applied because with 
${\mit\Sigma} /(2L) \approx \rho$ equation (\ref{tb}) is reduced to 
equation (\ref{tbequil}).  However, equation (\ref{bcr}) should be used 
for $B_{\rm cr}$ instead of equation (\ref{bcra}), which holds only for 
clouds in force balance along field lines.

The time scale $t_B$ given by equation (\ref{tbequil}) depends on the 
density and the magnetic field strength of the cloud, but does not depend 
on the cloud mass.  The gas temperature $T$ affects $t_B$ through densities 
of charged particles, $\tau_\lambda$, and $B_{\rm cr}$.  We take 
$T = 10$\,K in this paper.  

Figure 3 shows $t_B$ for the cases of $B = B_{\rm cr}$ (solid lines) and 
$B = 0.1 B_{\rm cr}$ (dashed lines) for clouds in force balance along field 
lines, or with $B_{\rm cr}$ given by equation (\ref{bcra}).  The dot-dashed 
lines show the Ohmic dissipation time $t_{\rm od}$, which is obtained by 
taking a limit $B \to 0$ in equation (\ref{tbequil}), or by setting 
$L \approx Z$ in equation (\ref{tbod}).  A relation 
$t_B = t_{\rm od} /(1 + D)$ holds, where $D$ is the excess dissipation 
relative to Ohmic dissipation given by equation (\ref{excess}).  
In addition to the {\it standard case} 
$\zeta_0 = 1 \times 10^{-17}\,{\rm s}^{-1}$ (thick lines), we also show 
the case of $\zeta_0 = 1 \times 10^{-16}\,{\rm s}^{-1}$ (thin lines) 
because there are some suggestions from observations of molecular ions 
that $\zeta_0$ might be significantly larger than the standard value in 
some clouds (de Boisanger, Helmich, \& van Dishoeck 1996; \cite{cas98}).  
At $n_{\rm H} \lesssim 10^7\,{\rm cm}^{-3}$, $t_B$ is 
10 to $10^2$ times $t_{\rm ff}$ for the case of $B=B_{\rm cr}$, and 
$t_B$ for the case of $B=0.1B_{\rm cr}$ is about $10^2$ times larger than 
that for $B=B_{\rm cr}$.  At $n_{\rm H} \gtrsim 10^8\,{\rm cm}^{-3}$, 
the ratio $t_B /t_{\rm ff}$ decreases as the density increases, and 
at last $t_B = t_{\rm ff}$ is attained at some density, which we denote 
$n_{\rm dec}$ and call the decoupling density as in our previous work.  
For the standard case $\zeta_0 = 1 \times 10^{-17}\,{\rm s}^{-1}$ we find 
$n_{\rm dec} \approx 2 \times 10^{11}$ and $3 \times 10^{11}\,{\rm cm}^{-3}$ 
for $B=B_{\rm cr}$ and $0.1B_{\rm cr}$, respectively.  
As shown by \cite{nis91}, $t_B$ and $n_{\rm dec}$ do not depend sensitively 
on various parameters such as fractions of heavy elements remaining 
in the gas phase and on the details of the grain model (e.g., 
$a_{\rm min}$).  As seen from Figure 3, $t_B$ and $n_{\rm dec}$ are not 
very sensitive to $\zeta_0$.

Ciolek \& Mouschovias (1993) criticized our previous work (e.g., 
\cite{nak86}; \cite{ume90}; \cite{nis91}) in their \S\,1.1 claiming that 
comparison of $t_B$ with $t_{\rm ff}$ and comparison of $v_B$ with the 
free-fall velocity $u_{\rm ff}$ were meaningless because magnetically 
subcritical clouds did not contract freely.  However, because the free-fall 
time is one of the fundamental time scales of clouds, we can find out by 
comparing $t_B$ with $t_{\rm ff}$ (or $v_B$ with $u_{\rm ff}$) whether 
the magnetic flux is lost effectively in dynamically contracting clouds, 
and how slowly the clouds in quasi-equilibrium contract induced by the 
drift of magnetic fields without detailed simulations which have been 
done by many authors (e.g., \cite{nak79}; \cite{liz89}; \cite{cio94}).  
According to Ciolek \& Mouschovias (1993), we concluded erroneously that 
ambipolar diffusion (or drift of magnetic fields) was inefficient at 
$n_{\rm H} \ll n_{\rm dec}$ because $t_B \gg t_{\rm ff}$.  Our conclusion 
in the previous and this papers is that extensive flux loss occurs only at 
$n_{\rm H} \gtrsim n_{\rm dec}$.  If there exist highly magnetically 
subcritical clouds with ${\mit\Phi} \gg {\mit\Phi}_{\rm cr}$, they might 
lose magnetic fluxes extensively even at $n_{\rm H} \ll n_{\rm dec}$ 
contracting quasistatically, though only down to 
$\approx {\mit\Phi}_{\rm cr}$, as shown by numerical simulations of 
Ciolek \& Mouschovias (1994).  However, clouds or cloud cores with 
${\mit\Phi} \gg {\mit\Phi}_{\rm cr}$ cannot exist (\cite{nak98}).  
Moreover, as mentioned in \S\,1, cloud cores must decrease their magnetic 
fluxes down to $10^{-3}\,{\mit\Phi}_{\rm cr}$ or below in the process of 
star formation.  Cloud cores with ${\mit\Phi}$ somewhat smaller than 
${\mit\Phi}_{\rm cr}$ can begin dynamical contraction rather easily 
(\cite{nak98}).  Extensive flux loss down to 
${\mit\Phi} \ll {\mit\Phi}_{\rm cr}$  does not occur at 
$n_{\rm H} \ll n_{\rm dec}$  because $t_B$ is much larger than the 
dynamical time scale.

\subsection{Dependence on Magnetic Field Strength} \label{bdependence}

We try to approximate the dependence of $t_B$ on the field strength $B$ 
by a power law $t_B \propto B^{-\gamma}$.  Figure 4 ({\it top}) shows 
the power index $\gamma$ obtained by comparing $t_B$ for the two cases 
$B = B_{\rm cr}$ and $0.1 B_{\rm cr}$ for the standard case 
$\zeta_0 = 1 \times 10^{-17}\,{\rm s}^{-1}$.   
At $n_{\rm H} \lesssim 10^7\,{\rm cm}^{-3}$, deviation of 
$\gamma$ from $2$ is small, or $t_B \propto B^{-2}$ approximately holds, 
at least for $B_{\rm cr} \geq B \gtrsim 0.1B_{\rm cr}$, in good agreement 
with ambipolar diffusion described in \S\,\ref{limitingcase}.  This is 
because ions and smallest grains, which are major contributors to 
transmitting the magnetic force to the neutrals, are relatively well 
frozen to magnetic fields in this density range.  We shall discuss the 
small deviation of $\gamma$ from 2 in \S\,\ref{contgrain}.

At $n_{\rm H} > 10^7\,{\rm cm}^{-3}$, $\gamma$ decreases steeply as the 
density increases, and finally settles down to $\approx 0$.  The situation 
$\gamma \ll 1$ suggests that Ohmic dissipation contributes at least 
significantly to magnetic flux loss.  However, although $\gamma = 0.13$ 
at $n_{\rm H} = 2 \times 10^{11}\,{\rm cm}^{-3}$, the flux loss occurs 
much faster than by Ohmic dissipation;  $t_B \approx 0.1 t_{\rm od}$ 
as shown in Figure 3, or $D \approx 10$.  The steep decrease of $\gamma$ 
with density and this behavior of magnetic flux loss at 
$n_{\rm H} \sim n_{\rm dec}$ can be understood by checking the motion of 
some typical charged particles.

Figure 4 ({\it bottom}) shows the drift velocity relative to that of 
magnetic fields, $v_{\rm i}/v_B$, given by equation (\ref{vlvb}), and 
$\tau_{\rm i}\omega_{\rm i}$ of the dominant ions M$^+$ for the two cases 
$B = B_{\rm cr}$ (solid line) and $0.1 B_{\rm cr}$ (dashed line) with 
$\zeta_0 = 1 \times 10^{-17}\,{\rm s}^{-1}$.  Here we omit the subscript 
$x$ to the velocities.  At $n_{\rm H} = 1 \times 10^9\,{\rm cm}^{-3}$, 
for example, metallic ions drift almost with magnetic fields with 
$v_{\rm i}/v_B \approx 0.96$ for $B=B_{\rm cr}$ because they are well 
frozen to magnetic fields with $\tau_{\rm i}\omega_{\rm i} \approx 32$.  
For $B=0.1B_{\rm cr}$, however, we find $v_{\rm i}/v_B \approx 0.26$ and 
$\tau_{\rm i}\omega_{\rm i} \approx 3.2$.  This value of $v_{\rm i}/v_B$ 
is much smaller than a naive estimation 
$v_{\rm i}/v_B \approx (\tau_{\rm i}\omega_{\rm i})^2 / 
   [1+(\tau_{\rm i}\omega_{\rm i})^2] \approx 0.91$ 
for this value of $\tau_{\rm i}\omega_{\rm i}$.  This large deviation 
is caused by the second term in the parentheses of equation (\ref{vlvb}), 
or by interaction with other charged particles, especially grains.  
Even for $B = B_{\rm cr}$ deviation of $v_{\rm i}$ from $v_B$ is caused 
mostly by this term:  a naive estimation gives 
   $v_{\rm i}/v_B \approx (\tau_{\rm i}\omega_{\rm i})^2 / 
                      [1+(\tau_{\rm i}\omega_{\rm i})^2] \approx 0.999$.  
As $B$ decreases, $v_{\rm i}/v_B$ decreases markedly.

The situation is the same for grains because 
$|\tau_{\rm g} \omega_{\rm g}| \ll \tau_{\rm i} \omega_{\rm i}$.  
Figure 5 shows the drift velocity relative to that of magnetic fields, 
$v_{\rm g}/v_B$, and $|\tau_{\rm g}\omega_{\rm g}|$ of grains of charge 
$-e$ as functions of the grain radius at several densities for 
$\zeta_0 = 1 \times 10^{-17}\,{\rm s}^{-1}$.  
At $n_{\rm H} = 10^9\,{\rm cm}^{-3}$ and $B = B_{\rm cr}$, grains of 
$a = 10$\,nm with electric charge $-e$ and $+e$, for example, have 
relative drift velocities $v_{\rm g}/v_B = 0.41$ and $-0.22$, respectively.  
These particles have $|\tau_{\rm g}\omega_{\rm g}| = 0.32$, and then the 
naive estimation gives   $v_{\rm g}/v_B \approx 
       (\tau_{\rm g}\omega_{\rm g})^2 / [1+(\tau_{\rm g}\omega_{\rm g})^2] 
    \approx 0.093$ 
for both charges, much smaller than the actual values; the actual velocity 
of $+e$ grains has even an opposite sign.  As $B$ decreases, 
$|v_{\rm g}/v_B|$ decreases markedly for all radii as shown in Figure 5.

At $n_{\rm H} > 10^7\,{\rm cm}^{-3}$  ions are not very well frozen to 
magnetic fields even for $B \approx B_{\rm cr}$.  Therefore, as $B$ 
decreases, the degree of freezing and $v_{\rm i}/v_B$ decrease markedly.  
The situation is the same for grains which are much less tightly coupled 
with magnetic fields.  Decrease of $v_{\rm i}/v_B$ and $|v_{\rm g}/v_B|$ 
has an effect of decreasing $t_B$ if $v_{\rm i}$ and $v_{\rm g}$ are fixed.  
On the other hand, as $B$ decreases, the magnetic force, or the driving 
force of the drift, decreases, which has an effect of decreasing the drift 
velocity $v_{\rm i}$ and $|v_{\rm g}|$ as seen from equation 
(\ref{frictionx}), and then has an effect of increasing $t_B$ if 
$v_{\rm i}/v_B$ and $v_{\rm g}/v_B$ are fixed.  These opposite effects 
make $t_B$ less sensitive to $B$ than at $n_{\rm H} < 10^7\,{\rm cm}^{-3}$, 
where ions and smallest grains, the dominant charged particles, are 
relatively well frozen and the decrease of $B$ has little effect on their 
$v_\lambda/v_B$.  As the density increases, $t_B$ becomes more and more 
insensitive to $B$ because $|\tau_\lambda \omega_\lambda|$  decreases with 
density for all charged particles even though $t_B$ is significantly 
smaller than $t_{\rm od}$.  This is why $\gamma$ decreases steeply with 
density at $n_{\rm H} > 1 \times 10^7\,{\rm cm}^{-3}$.

\subsection{Contribution of Grains} \label{contgrain}

Even at $n_{\rm H} < 1 \times 10^7\,{\rm cm}^{-3}$, $\gamma$ shows some 
deviation from 2 as seen in Figure 4 ({\it top}).  Because ions are 
strongly coupled with magnetic fields and deviation of $v_{\rm i}$ from 
$v_B$ is very small at these densities as shown in Figure 4 ({\it bottom}), 
this deviation suggests that transmission of the magnetic force to the 
neutrals is significantly contributed by grains, which are not very 
strongly coupled with magnetic fields even at these low densities as shown 
in Figure 5.

To confirm this we calculate the frictional force exerted by each kind of 
charged particles on the neutrals, which is given by each term of equation 
(\ref{frictionx}).  We shall not consider the component of the frictional 
force perpendicular to both $\mbf{B}$ and $\mbf{j \times B}$ because 
summation of this component for all charged particles vanishes as shown 
by equation (\ref{frictiony}).  Figure 6 shows the frictional forces of 
ions and grains relative to the total frictional force, which is equal to 
$|\,\mbf{j \times B}|/c$ per unit volume, as functions of the cloud 
density for the two cases of the field strength with 
$\zeta_0 = 1 \times 10^{-17}\,{\rm s}^{-1}$.  Ions contribute more than 
grains only at  $n_{\rm H} \lesssim 10^4 \,{\rm cm}^{-3}$, where the 
ionization fraction is relatively high.  
At $n_{\rm H} \gtrsim 10^6 \,{\rm cm}^{-3}$  grains contribute more than 
about 90\% of the frictional force at least for 
$B_{\rm cr} \geq B \gtrsim 0.1B_{\rm cr}$.  The frictional force exerted 
by electrons is more than 3 orders of magnitude smaller than that by ions 
mainly because of their small drift momentum.  For the case of 
$\zeta_0 = 1 \times 10^{-16}\,{\rm s}^{-1}$ these densities are somewhat 
higher;  e.g., ions contribute more than grains at 
$n_{\rm H} \lesssim 10^5\,{\rm cm}^{-3}$ because of higher ion densities.

Figures 7 and 8 show the frictional force exerted by grains on the 
neutrals as a function of their radius $a$ and charge $q_{\rm g}$ at 
several cloud densities for the two cases of the field strength 
$B = B_{\rm cr}$ and $0.1 B_{\rm cr}$ with 
$\zeta_0 = 1 \times 10^{-17}\,{\rm s}^{-1}$.  As well as ions, negatively 
charged grains exert frictional force, or drift, in the direction of the 
magnetic force irrespective of their radius at 
$n_{\rm H} \lesssim 10^{14}(B/B_{\rm cr})^2{\rm cm}^{-3}$ at least for 
$B \gtrsim 0.1 B_{\rm cr}$.  Positively charged grains behave differently.  
At low densities they contribute in the same direction as negatively 
charged grains irrespective of their radius.  However, at some density the 
largest grains begin to contribute, or drift, in the opposite direction, 
and the size range of such grains expands to smaller radii as $n_{\rm H}$ 
increases; frictional forces in these ranges are shown by dotted lines in 
Figures 7 and 8.  For example, grains of charge $+e$ with $a > 160$\,nm 
drift in the opposite direction at $n_{\rm H} = 10^5\,{\rm cm}^{-3}$ for 
$B = B_{\rm cr}$, and those with  $a > 93$\,nm do at 
$n_{\rm H} = 10^4\,{\rm cm}^{-3}$ for $B = 0.1 B_{\rm cr}$ though the 
dotted line is outside this panel.  Finally at 
$n_{\rm H} \approx 2 \times 10^9$ and $9 \times 10^6 \,{\rm cm}^{-3}$ 
for $B=B_{\rm cr}$ and $0.1B_{\rm cr}$, respectively, even the smallest 
grains ($a \approx 5$\,nm) with charge $+e$ begin to drift opposite to 
the magnetic force.  These are because $A_2 <0$ (or $\sigma_{\rm H} < 0$) 
in most of the density region covered by this paper (see Figure 9; 
$A_1 > 0$ by definition), and thus positively charged particles with small 
$\tau_\lambda \omega_\lambda$ have negative drift velocities as seen from 
equation (\ref{vlvb}).  Because 
$\tau_{\rm g} \omega_{\rm g} \propto a^{-2} n_{\rm H}^{-1/2}(B/B_{\rm cr})$, 
the radius range in which positively charged grains have negative drift 
velocities expands to smaller $a$ as $n_{\rm H}$ increases for a fixed 
$B/B_{\rm cr}$ as long as $A_2 < 0$.  We shall discuss the physical reason 
of this phenomenon in \S\,\ref{driftgrain}.

At high densities, not only $t_B$ deviates greatly from the 
$t_B \propto B^{-2}$ relation, but also grains, the main contributors to 
the frictional force, of opposite charges drift in opposite directions.  
This contradicts the literal meaning of ambipolar diffusion, a term 
originally used in plasma physics for a quite different phenomenon in 
which particles of opposite charges diffuse as one unit (e.g., 
\cite{cap76}; see also Appendix~\ref{termin}).

The great contribution of grains to the frictional force comes from 
their large contribution to $A_1$ and $A_2$, or $\sigma_{\rm P}$ and 
$\sigma_{\rm H}$, though $\sigma_\parallel$ is mainly contributed by 
electrons in most of the density region in this paper.  This difference 
in the contribution is due to large differences in 
$|\tau_\lambda \omega_\lambda|$ and in the abundance $n_\lambda /n_{\rm H}$ 
among charged particles.  For metallic ions we have  
    $\tau_{\rm i} \omega_{\rm i} \approx 
            320 (n_{\rm H} /10^7\,{\rm cm}^{-3})^{-1/2}(B/B_{\rm cr})$.  
At $T \approx 10$\,K we find
    $|\tau_{\rm e}\omega_{\rm e}| / \tau_{\rm i} \omega_{\rm i} 
            \approx 4.8 \times 10^3$   and 
    $|\tau_{\rm g}\omega_{\rm g}| / \tau_{\rm i}\omega_{\rm i} 
            \approx 1.0 \times 10^{-2}(a/10\,{\rm nm})^{-2}$  
for grains of charge $q_{\rm g}=\pm e$.  
For the case of $\zeta_0 = 1 \times 10^{-17}\,{\rm s}^{-1}$, 
$\sigma_\parallel$ is mainly contributed by electrons at 
$n_{\rm H} \lesssim 10^{13}\,{\rm cm}^{-3}$ and by grains at higher 
densities as shown in Figure 9.  Contribution of ions to $\sigma_\parallel$ 
is minor at all densities.  To $\sigma_{\rm P}$ or $A_1$, ions contribute 
more than half only at $n_{\rm H} \lesssim 10^4 \,{\rm cm}^{-3}$, and grains 
contribute more than 80\% at $n_{\rm H} \gtrsim 10^5 \,{\rm cm}^{-3}$ for 
$B = B_{\rm cr}$ (Figure 9), though for $B = 0.1 B_{\rm cr}$ ions also 
contribute $50 - 80\%$ at $n_{\rm H}$ between $3 \times 10^8$ and  
$6 \times 10^{10}\,{\rm cm}^{-3}$.  
At $n_{\rm H} \gtrsim 10^{11}\,{\rm cm}^{-3}$, grains overwhelm the other 
particles.  Contribution of electrons to $\sigma_{\rm P}$ is less than 
$10^{-3}$ because of large $|\tau_{\rm e}\omega_{\rm e}|$ at low densities 
and because of low abundance at high densities [see equation (\ref{a1}) or 
(\ref{sigmap})].  Situation is complicated for $\sigma_{\rm H}$ or $A_2$.  
Because the terms of some given particles have quite different forms even 
with opposite signs between the equivalent equations (\ref{a2}) and 
(\ref{sigmah}), we cannot uniquely tell how much each kind of particles 
contributes to $\sigma_{\rm H}$.  Moreover, particles of opposite charges 
contribute oppositely and significant cancellation occurs in some 
situations for either expression of $\sigma_{\rm H}$.  We would make 
a serious mistake if we neglect the terms of grains in $\sigma_{\rm H}$ 
except at high densities where both 
$\tau_{\rm i} \omega_{\rm i} \lesssim 10$ and $n_{\rm i} \gg n_{\rm e}$ hold.

The sign of $\sigma_{\rm H}$ changes at some density as shown in Figure 9.  
With an expression of $\sigma_{\rm H}$ accurate at very high densities, 
we find that this occurs when 
    $\tau_{\rm i} \omega_{\rm i} \approx (n_{\rm e}/n_{\rm i})^{1/2}$.  
As long as $B \gtrsim 0.01B_{\rm cr}$, this gives 
$\tau_{\rm i}\omega_{\rm i} \approx 0.08$, which corresponds to the density 
$n_{\rm H} \approx 1.6 \times 10^{14} (B/B_{\rm cr})^2\,{\rm cm}^{-3}$.  
Because of the change of the sign, grains change the directions of drift 
at densities somewhat higher than this as seen from equation (\ref{vlvb}).  
However, the total frictional force they exert on the neutrals is always 
in the direction of magnetic force because the mean drift velocity of $-e$ 
and $+e$ grains is in the direction of magnetic force as seen from equation 
(\ref{vlx}) or (\ref{vlvb}).  

Desch \& Mouschovias (2001) ignored the contribution of grains to 
$\sigma_{\rm P}$ ($\sigma_\perp$ by their notation) and $\sigma_{\rm H}$ 
as seen from their equations (25) and (26).  Therefore, their equation 
(28), which was obtained by using these equations, does not correspond 
to the critical state at which Ohmic dissipation becomes important, 
or $D \approx 1$, contrary to their statement.  Their results will be 
discussed as compared with ours in \S\,\ref{comparison}.

Because the frictional force is contributed mainly by grains, $t_B$ is 
determined mainly by grains except at very low densities.  Besides, the 
reactions at grain surface are important in determining the densities of 
various charged particles (e.g., \cite{nak84}; \cite{nis91}).  Thus, grains 
play a decisive role in the process of magnetic flux loss in molecular 
clouds.

\section{Discussion}

\subsection{Drift of Grains} \label{driftgrain}

Grains of opposite electric charges drift in opposite directions at high 
densities as shown in \S\S\,\ref{bdependence} and \ref{contgrain}.  This is 
formally a result of the second term in the parentheses of equation 
(\ref{vlvb}).  Here we shall give a more intuitive explanation of this 
phenomenon.

In the existence of an electric field  $\mbf{E}$  a particle of electric 
charge $q_\lambda \neq 0$ gyrating around magnetic field lines\footnote{
    One may think that gyration completely dissipates because the viscous 
    damping times $\tau_\lambda$ are smaller than $t_B$ and $t_{\rm ff}$ 
    by orders of magnitude.  However, because of collisions with neutral 
    molecules and atoms charged particles always have thermal motions, 
    which would cause gyration, though a term "gyrating" may not be 
    appropriate when $|\tau_\lambda \omega_\lambda| \lesssim 1$.} 
drifts with a velocity  $\mbf{v}_\lambda = c\,\mbf{E \times B}/B^2$  
independent of its charge.  If there is a force field  $\mbf{f}$, which is 
independent of the particle charge, instead of the electric field, the 
particle drifts with a velocity  
     $\mbf{v}_\lambda = (c/q_\lambda)\,\mbf{f \times B}/B^2$  
dependent on its charge.  In the frame moving with the neutrals in 
a molecular cloud, there is an electric field satisfying equation 
(\ref{elecfield}), which causes a charge-independent drift motion (call 
this drift 1).  Neutral molecules exert a frictional force on this drift 
motion, which is anti-parallel to the drift velocity independent of the 
electric charge.  Therefore, this force causes another drift motion 
(drift 2) whose direction depends on the particle charge.  Thus the drift 
velocity has both components dependent on and independent of the particle 
charge.  Because drift 2 has opposite directions for particles of opposite 
charges, the frictional force on this motion yields the third drift motion 
whose direction is charge-independent, etc.

All these can be taken into account in the equation for the steady motion 
(\ref{meanmotion}) in Appendix~\ref{anotherform}.  The direct solution of 
this equation is given by equations (\ref{vlambdapara}) and 
(\ref{vlambdaperp}), whose components perpendicular to  $\mbf{B}$  are 
found to agree with equations (\ref{vlx}) and (\ref{vly}).  When 
     $|\tau_\lambda \omega_\lambda| \lesssim 1$, 
the frictional force can be comparable to or stronger than the electric 
force, and therefore the charge-dependent part of the drift velocity can 
be comparable to or greater than the charge-independent one.

Because grains of opposite charges drift in opposite directions at high 
densities where grains are the major charged particles, one may anticipate 
that charge separation occurs in the $\mbf{j \times B}$\,-, or 
$x$-direction.  However, this is not the case.  All charged particles 
should be taken into account.  As confirmed in \S\,\ref{formulae}, 
equation (\ref{vlx}) guarantees that the $x$-component of the electric 
current density vanishes.

Although the frictional force is contributed mostly by grains at high 
densities, their drift velocities are very small.  For example, at the 
decoupling density 
$n_{\rm H} \approx n_{\rm dec} \approx 2 \times 10^{11}\,{\rm cm}^{-3}$ 
for $B = B_{\rm cr}$ with $\zeta_0 = 1 \times 10^{-17}\,{\rm s}^{-1}$ 
even the smallest grains of $a \approx 5$\,nm with electric charge $-e$ 
have a drift velocity as small as $v_{\rm g}/v_B \approx 0.02$ as seen 
from Figure 5.  Because extensive magnetic flux loss occurs only at 
$n_{\rm H} \gtrsim n_{\rm dec}$ and decrease of $B/B_{\rm cr}$ by the flux 
loss makes $v_{\rm g}/v_B$ even smaller (Figure 5), grains are hardly lost 
from the cloud core even if the magnetic flux decreases by orders of 
magnitude as far as the cloud core is magnetically supercritical.  
Although Ciolek \& Mouschovias (1994) write that the grain abundance 
decreases almost in proportion to the reduction factor of the magnetic 
flux in highly magnetically subcritical cloud cores, cloud cores can 
hardly be magnetically subcritical as discussed by Nakano (1998).  
Although the drift velocity of ions is not very small compared with that 
of magnetic fields even at $n_{\rm H} \approx n_{\rm dec}$ for 
$B \approx B_{\rm cr}$ (Figure 4), ions are quite minor constituents 
among the heavy elements even in the gas phase (Figure 1).  Thus magnetic 
flux loss in cloud cores has little effect on the abundance of heavy 
elements in stars born therein.

\subsection{Charge Fluctuation of Grains} \label{cfluct}

Electric charges of grains change stochastically because ions and 
electrons sometimes stick to them.  This change has some effect on their 
motion in electromagnetic fields.  Nakano (1984) and \cite{nak86} 
neglected this effect because they had a rough estimation that it was 
small (see also \cite{nak80}).

Kamaya \& Nishi (2000) investigated this effect more elaborately for 
grains of charge $-e$.  The probability of having some charge is 
proportional to the time span during which the grain has this charge.  
Therefore, as seen from Figure 2, the grains of charge $-e$ change their 
charge mostly by going to the neutral state at most radii $a$ and at most 
densities $n_{\rm H}$.  The charge change between the $-e$ and neutral 
states effectively decreases the viscous damping time $\tau_{\rm g}$ of 
grains of charge $-e$ by some factor $C_{\rm g} > 1$.  This effect can be 
taken into account by replacing $\tau_{\rm g}$ with 
$\tau_{\rm g}/C_{\rm g}$ in our formalism summarized in \S\,\ref{drift}.  
Kamaya \& Nishi found that if H$_3^+$ is the dominant ion, $C_{\rm g}$ 
for grains of $a = 10$\,nm is 1.3 at 
$n_{\rm H} \approx 10^3 \,{\rm cm}^{-3}$ and approaches $1$ as $n_{\rm H}$ 
increases, and that larger grains have $C_{\rm g}$ closer to $1$.

Because metallic ions and molecular ions other than H$_3^+$ are the 
dominant ions as shown in Figure 1, deviations of $C_{\rm g}$ from $1$ 
are about 3 times smaller than those obtained for H$_3^+$ by Kamaya \& 
Nishi (2000) at all densities and at all grain radii;  the deviation 
$C_{\rm g} - 1$ is proportional to $m_{\rm i}^{-1/2}$, where $m_{\rm i}$ 
is the mass of the dominant ions.  Therefore, charge fluctuation of 
grains can hardly affect $t_B$ because $C_{\rm g}$ can be as large as 
1.1 only for smallest grains only at low densities 
$n _{\rm H} \sim 10^3 \,{\rm cm}^{-3}$, where grains are minor 
contributors to the frictional force.  Thus, our formalism can be applied 
to any situation with little error though Ciolek \& Mouschovias (1993) 
pointed out our ignorance of this effect.

\subsection{Comparison with Other Work} \label{comparison}

We compare our results with several previous works on $t_B$, 
$n_{\rm dec}$, and the dissipation in excess of Ohmic dissipation given 
by $D$ in equation (\ref{excess}).

The decoupling densities $n_{\rm dec}$ obtained in \S\,\ref{losstime} are 
about 4 times larger than that of \cite{nis91} for the same grain model.  
The main cause for this difference is in $B_{\rm cr}$;  our $B_{\rm cr}$ 
is $\sqrt 3$ times larger than theirs.  At relatively low densities where 
major charged particles are well frozen to magnetic fields, difference 
in the coefficient of $B_{\rm cr}$ has little effect on $t_B$ for a given 
$B/B_{\rm cr}$ because these particles drift almost with magnetic fields 
and the drift velocity $v_{Bx}$ is determined by the balance of the 
frictional force on them with $(B/B_{\rm cr})^2$ times the 
gravity as seen from equations (\ref{vbambi}) and (\ref{magforce}).  
At high densities, $t_B$, and then $n_{\rm dec}$, depend on the coefficient 
of $B_{\rm cr}$ because the delay in the drift of major charged 
particles, which are not well frozen, depends on the field strength as 
discussed in \S\,\ref{bdependence}.  However, the general feature of 
$t_B$ (rough dependences on $n_{\rm H}$ and $B /B_{\rm cr}$, existence of 
$n_{\rm dec}$, etc.) is hardly affected by this coefficient.

\cite{nak86} investigated magnetic flux loss in spherical clouds, and 
found that magnetic decoupling occurs at 
$n_{\rm H} \approx n_{\rm dec} \approx 5 \times 10^{11}\,{\rm cm}^{-3}$ 
for a cloud of $1M_\odot$ and the decoupling density does not depend 
sensitively on the cloud mass.  They also found that dissipation in excess 
of Ohmic one is not large, or $D \sim 1$, at $n_{\rm H} \approx n_{\rm dec}$ 
for $B \approx B_{\rm cr}$.

In \S\,\ref{bdependence} we have found $D \approx 10$ at 
$n_{\rm H} \approx n_{\rm dec}$, which is an order of magnitude larger 
than $D$ obtained by \cite{nak86}.  We can list up some causes for this 
discrepancy;  differences in the cloud model and in the grain model.  
\cite{nak86} assumed that clouds were spherical without force balance 
along field lines.  They also assumed for simplicity that all grains had 
the same radius $a = 100$\,nm.  Another difference is in the electric 
conductivity $\sigma_\parallel$.  For the collision of electrons with 
the neutrals, \cite{nak86} adopted the classical Langevin's cross sections 
in which electric polarization of the neutrals induced by an approaching 
electron has a great effect.  These were widely adopted formerly 
(e.g., \cite{nak84}; \cite{mou91}) assuming the similarity with the 
collision of ions with the neutrals.  However, we found that laboratory 
experiments show much smaller cross sections at low energies 
(\cite{hay81}), which are not much different from the geometrical cross 
sections.  Sano et al. (2000) give the empirical formulae of the momentum 
transfer rate coefficients for the collision of electrons with H$_2$ 
molecules and with He atoms, which were obtained by Umebayashi (1993, 
private communication) by fitting to the laboratory data.  With these 
collision rate coefficients, which are about 80 times smaller than the 
previous ones at $T \approx 10\,$K, the conductivity $\sigma_\parallel$, 
$t_{\rm od}$, and $D$ are much larger than the previous ones except at 
$n_{\rm H} \gtrsim 10^{13}\,{\rm cm}^{-3}$ where $\sigma_\parallel$ is 
mainly contributed by grains.  On the other hand, $\sigma_{\rm P}$ is 
hardly affected by this change of the cross sections because contribution 
of electrons is very small even with the enhanced $\sigma_{\rm e}$ 
(Figure 9), and $\sigma_{\rm H}$ is also hardly affected except at very 
high densities where $|\sigma_{\rm H}| \ll \sigma_{\rm P}$.  Therefore, 
$t_B = t_{\rm od}/(1 + D)$ is hardly affected by this change because 
$t_{\rm od} \propto \sigma_\parallel$ and 
        $1 + D = \sigma_\parallel \sigma_{\rm P}
                     /(\sigma_{\rm P}^2 + \sigma_{\rm H}^2)$.

Desch \& Mouschovias (2001) write that Ohmic dissipation becomes important, 
or $D \approx 1$ is realized, when 
$|\tau_{\rm e} \omega_{\rm e}| \approx 6$.  However, 
$|\tau_{\rm e} \omega_{\rm e}| \approx 6$ was obtained from their 
equation (28),  $\tau_{\rm i} \omega_{\rm i}|\tau_{\rm e} \omega_{\rm e}|
                       \approx n_{\rm e}/n_{\rm i}$,\footnote{
   Emphasizing the contrast with this equation, Desch \& Mouschovias (2001) 
   write that \cite{nak86} used a relation 
     $\tau_{\rm i} \omega_{\rm i}|\tau_{\rm e} \omega_{\rm e}| \approx 1$ 
   for $D \approx 1$ assuming $n_{\rm e} = n_{\rm i}$ even at densities 
   as high as $n_{\rm H} \gtrsim n_{\rm dec}$.  This is another 
   misrepresentation of the fact in addition to their comments on 
   \protect\cite{nis91} cited in footnote 2.  Although \protect\cite{nak86} 
   gave the expressions of $D$, by their equations (49) and (50), when ions 
   and electrons are major contributors to $A_1$ and $A_2$ 
   (or $\sigma_{\rm P}$ and $\sigma_{\rm H}$) 
   (of course $n_{\rm e} \approx n_{\rm i}$), and when grains of charge 
   $-e$ and $+e$ are major contributors, respectively, as limiting cases, 
   they used the general expression of $D$ given by their equation (48) or 
   ours (\protect\ref{excess}) in numerical calculation.  For instance, 
   to calculate $D$ shown in their figure 2, they used the densities of 
   charged particles shown in their figure 3, which clearly shows the 
   deviation of $n_{\rm e}$ from $n_{\rm i}$ at high densities in agreement 
   with our equation (\protect\ref{neni}).}  
which does not correspond to the state of $D \approx 1$ 
because they ignored the contribution of grains to $\sigma_{\rm P}$ and 
$\sigma_{\rm H}$ (see \S\,\ref{contgrain}).  Their criterion 
$|\tau_{\rm e} \omega_{\rm e}| \approx 6$ for $D \approx 1$ gives a density 
$n_{\rm H} \approx 6 \times 10^{17}(B/B_{\rm cr})^2\,{\rm cm}^{-3}$ 
irrespective of $\zeta_0$ when $B_{\rm cr}$ is given by equation 
(\ref{bcra}), or the pressure force balances with the gravity along field 
lines.  When contraction along field lines is insufficient and then the 
pressure force along field lines is weaker than the gravity, $\mit\Sigma$ 
is larger than given by equation (\ref{sigma}) and $B_{\rm cr}$ is 
larger than equation (\ref{bcra}) for a given density.  This means that 
$|\tau_{\rm e} \omega_{\rm e}|$ is larger for given $n_{\rm H}$ and 
$B / B_{\rm cr}$, and therefore $|\tau_{\rm e} \omega_{\rm e}| \approx 6$ 
gives even higher densities than in the above case.

Taking into account the contribution of grains to $\sigma_{\rm P}$ and 
$\sigma_{\rm H}$, we have found that $D \approx 1$, or 
$t_B \approx t_{\rm od}/2$, is realized at 
$n_{\rm H} \approx 1 \times 10^{13}\,{\rm cm}^{-3}$ 
for $\zeta_0 = 1 \times 10^{-17}\,{\rm s}^{-1}$ almost independent of 
$B/B_{\rm cr}$ at least at $0.1 \lesssim B/B_{\rm cr} \le 1$.  This can 
be confirmed by substituting the values of $\sigma_\parallel$, 
$\sigma_{\rm P}$, and $\sigma_{\rm H}$ shown in Figure 9 into equation 
(\ref{excess}) and by noting that $t_B$ is almost independent of 
$B /B_{\rm cr}$ at high densities as shown in Figure 3.  This density is 
lower than Desch \& Mouschovias' values by orders of magnitude.

In reality, however, Ohmic dissipation becomes important at significantly 
lower densities.  At $n_{\rm H} \gtrsim n_{\rm dec}$ the clouds contract 
dynamically adjusting $B /B_{\rm cr}$ so as to keep 
$t_B \approx t_{\rm ff}$ because too much decrease of $B /B_{\rm cr}$ 
makes $t_B$ larger than $t_{\rm ff}$ as long as $t_{\rm od} > t_{\rm ff}$.  
However, $t_B = t_{\rm od}/(1 + D)$ cannot be larger than $t_{\rm od}$.  
Therefore, after $t_{\rm od} \approx t_{\rm ff}$ is attained at 
$n_{\rm H} \approx 1 \times 10^{12}\,{\rm cm}^{-3}$ (Figure 3), field 
dissipation proceeds by Ohmic dissipation faster than the contraction.

Kamaya \& Nishi (2000) investigated the motion of ions and grains in the 
process of magnetic flux loss using a simplified model.  Assuming that all 
grains have the same radius and have an electric charge $-e$ or 0, ions 
are well frozen to magnetic fields ($\tau_{\rm i} \omega_{\rm i} \gg 1$), 
and electrons are completely frozen 
($|\tau_{\rm e} \omega_{\rm e}| \rightarrow \infty$), they obtained the 
drift velocities of ions and grains of charge $-e$.  If we apply equation 
(\ref{vlx}) to their model, we easily obtain the drift velocities given by 
their equations (34) and (35) though $\tau_{\rm g}$ should be replaced by 
$\tau_{\rm g}/C_{\rm g}$.  Thus, Kamaya \& Nishi (2000) obtained some 
relatively transparent results due to their simplified model, which are 
consistent with our formalism, though they got some results different from 
Ciolek \& Mouschovias (1993).

\section{Summary}

We analyzed the detailed processes operating in the drift of magnetic 
fields in molecular clouds taking into account charged grains with 
the MRN size distribution in addition to ions and electrons and the effect 
of partial freezing of these particles to magnetic fields using the 
formalism obtained by Nakano (1984) and \cite{nak86}.

We found that to the frictional force, whereby the magnetic force is 
transmitted to neutral molecules, ions contribute more than half only at 
cloud densities $n_{\rm H} \lesssim 10^4 \,{\rm cm}^{-3}$, and grains 
contribute more than about 90\% at $n_{\rm H} \gtrsim 10^6 \,{\rm cm}^{-3}$.  
Besides, the reactions at grain surface are important in determining the 
densities of various charged particles.  Thus grains play a decisive role 
in the process of magnetic flux loss in molecular clouds.

We confirmed the previous results (\cite{nis91}) on the magnetic flux loss 
time $t_B$ and the decoupling density $n_{\rm dec}$, at which $t_B$ is 
equal to the free-fall time $t_{\rm ff}$ and therefore magnetic fields 
are effectively decoupled from the gas; e.g., $t_B \gg t_{\rm ff}$ at 
$n_{\rm H} \ll n_{\rm dec}$, and $n_{\rm dec}$ is almost independent 
of the magnetic field strength in the cloud and takes a value a few 
$\times 10^{11}\,{\rm cm}^{-3}$.  We found that $t_B$ and $n_{\rm dec}$ 
are not very sensitive to the ionization rate by cosmic rays, $\zeta_0$.

We investigated the dependence of $t_B$ on the field strength $B$.  
Approximating the relation by a power law $t_B \propto B^{-\gamma}$, 
we found $\gamma \approx 2$, characteristic to ambipolar diffusion, 
only at $n_{\rm H} \lesssim 10^7 \,{\rm cm}^{-3}$ where ions and smallest 
grains are pretty well frozen to magnetic fields.  
At $n_{\rm H} > 10^7 \,{\rm cm}^{-3}$, $\gamma$ decreases steeply as 
$n_{\rm H}$ increases, and finally at $n_{\rm H} \approx n_{\rm dec}$, 
$\gamma \ll 1$ is attained, reminiscent of Ohmic dissipation, though 
the flux loss occurs about 10 times faster than by Ohmic dissipation 
at $n_{\rm H} \approx$ a few\,$\times 10^{11}\,{\rm cm}^{-3}$.  Because 
even ions are not very well frozen to magnetic fields at 
$n_{\rm H} > 10^7 \,{\rm cm}^{-3}$, ions and charged grains drift slower 
than the magnetic fields.  Decrease of $B$ has an effect of increasing 
this lag.  This insufficient freezing makes $t_B$ less sensitive to $B$ 
than at $n_{\rm H} \lesssim 10^7 \,{\rm cm}^{-3}$ where major charged 
particles are well frozen and the decrease of $B$ has little effect on 
their lag.  This tendency is enhanced as the density increases, and 
at last $t_B$ becomes almost independent of $B$ at 
$n_{\rm H} \approx n_{\rm dec}$.  Ohmic dissipation is dominant only at 
$n_{\rm H} \gtrsim 10^{12}\,{\rm cm}^{-3}$.

We found that while ions and electrons drift in the direction of magnetic 
force at all densities, grains of opposite charges drift in opposite 
directions at high densities, where grains are major contributors to 
the frictional force, apart from the component of the drift velocities 
perpendicular to the magnetic force which yields no net frictional force.  
Although magnetic flux loss occurs significantly faster than by Ohmic 
dissipation even at $n_{\rm H} \approx n_{\rm dec}$, the operating process 
at high densities is quite different from ambipolar diffusion in which 
particles of opposite charges are supposed to drift as one unit.

\acknowledgments

This work was supported in part by the Grant-in-Aid for Scientific Research 
of Priority Areas (A) (No. 10147101, 10147105) of the Ministry of Education, 
Culture, Sports, Science, and Technology of Japan.

\appendix

\section{Another Method of Formulation} \label{anotherform}

Because the procedures of obtaining the formulae summarized in 
\S\,\ref{formulae} are rather complicated, it would be worthwhile to show 
another method of obtaining them.

With the same approximation as adopted by Nakano (1984) and \cite{nak86} 
and described in \S\,\ref{formulae}, the mean motion of charged particle 
$\lambda$ obeys
\begin{equation}
   q_\lambda \bigg( \mbf{E} + \frac{1}{c}\,\mbf{v_\lambda \times B} \bigg)
         - \frac{m_\lambda \mbf{v}_\lambda}{\tau_\lambda} = 0,
                                  \label{meanmotion}
\end{equation}
in the frame moving with the neutrals (see \S\,\ref{formulae} for the 
notation).  The solution of this equation is given by
\begin{equation}
   \mbf{v}_{\lambda \parallel} = \frac{q_\lambda \tau_\lambda}{m_\lambda}
                         \mbf{E}_\parallel,    \label{vlambdapara}
\end{equation}
\begin{equation}
   \mbf{v}_{\lambda \perp}
            = \frac{c}{B} \, \frac{(\tau_\lambda \omega_\lambda)^2}
                                  {1 + (\tau_\lambda \omega_\lambda)^2}
    \bigg(\frac{\mbf{E}_\perp}{\tau_\lambda \omega_\lambda} 
      + \mbf{E_\perp \times}\frac{\mbf{B}}{B}\bigg),
                                 \label{vlambdaperp}
\end{equation}
where subscripts $\parallel$ and $\perp$ represent the components parallel 
and perpendicular to  $\mbf{B}$, respectively.

The electric current density \mbf{j} is obtained by summing up 
$n_\lambda q_\lambda \mbf{v}_\lambda$  for all kinds of charged 
particles using equations (\ref{vlambdapara}) and (\ref{vlambdaperp}).  
Moving to a more general frame wherein the neutrals move with a velocity 
$\mbf{u}_{\rm n}$  and the electric field is given by 
      $\mbf{E}_0 = \mbf{E} - \mbf{u_{\rm n} \times B}/c$, 
and using the electrical neutrality relation, we obtain
\begin{equation}
   \mbf{j} = \sigma_\parallel \mbf{E}_{0\parallel}
          + \sigma_{\rm P} \bigg(\mbf{E}_{0\perp}
                             + \frac{1}{c} \,\mbf{u_{\rm n}\times B}\bigg)
          + \sigma_{\rm H} \frac{\mbf{B}}{B} \mbf{\times}
                           \bigg(\mbf{E}_{0\perp}
                             + \frac{1}{c} \,\mbf{u_{\rm n} \times B}\bigg),
                                    \label{current}
\end{equation}
where $\sigma_\parallel$ is the electric conductivity along magnetic field 
lines given by equation (\ref{sigmapara}), and $\sigma_{\rm P}$ and 
$\sigma_{\rm H}$ are Pedersen and Hall conductivities, respectively, given by
\begin{equation}
   \sigma_{\rm P} = \sum_\nu \frac{\sigma_\nu}
                                  {1 + (\tau_\nu \omega_\nu)^2}
                  = \bigg(\frac{c}{B}\bigg)^2 A_1,    \label{sigmap}
\end{equation}
\begin{equation}
   \sigma_{\rm H} = -\sum_\nu \frac{\sigma_\nu \tau_\nu \omega_\nu}
                                   {1 + (\tau_\nu \omega_\nu)^2}
                  = \bigg(\frac{c}{B}\bigg)^2 A_2,    \label{sigmah}
\end{equation}
where $A_1$, $A_2$, and $\sigma_\nu$ are given by equations (\ref{a1}),  
(\ref{a2}), and (\ref{sigmapara}), respectively.  The last equality of 
equation (\ref{sigmah}) can be found by using the electrical neutrality 
relation.  Equation (\ref{current}) with (\ref{sigmapara}), (\ref{sigmap}), 
and (\ref{sigmah}) is a generalization of, e.g., Parks (1991) who 
considered only electrons and a single kind of ions as charged particles.

The drift velocity of magnetic fields, $\mbf{v}_B$, satisfies equation 
(\ref{elecfield}).  Making a vector product of equation (\ref{current}) 
with $\mbf{B}$ and eliminating 
     $\mbf{E}_0 + \mbf{u_{\rm n} \times B}/c = \mbf{E}$ 
by using equation (\ref{elecfield}), we obtain
\begin{equation}
   \frac{1}{c}\, \mbf{j \times B} = \bigg(\frac{B}{c}\bigg)^2
     \bigg(\sigma_{\rm P}\mbf{v}_B + \sigma_{\rm H} \frac{\mbf{B}}{B}
                                  \mbf{\times v}_B \bigg). \label{magforcea}
\end{equation}
This equation is solved for $\mbf{v}_B$ as
\begin{equation}
   \mbf{v}_B = \bigg(\frac{c}{B}\bigg)^2 \, \bigg[ \,
       \frac{\sigma_{\rm P}}{\sigma_{\rm P}^2 + \sigma_{\rm H}^2} \,
          \frac{1}{c}\, \mbf{j \times B}
     + \frac{\sigma_{\rm H}}{\sigma_{\rm P}^2 + \sigma_{\rm H}^2} \,
          \frac{1}{c} \,(\mbf{j \times B) \times} \frac{\mbf{B}}{B} \,\bigg].
                          \label{vbvec}
\end{equation}
By taking the components of this equation we can confirm that equation 
(\ref{vbvec}) is identical with equations (\ref{vbx}) and (\ref{vby}).

Eliminating $\mbf{E}_\perp$ in equation (\ref{vlambdaperp}) by using 
equation (\ref{elecfield}), we obtain
\begin{equation}
   \mbf{v}_{\lambda\perp} = \frac{(\tau_\lambda \omega_\lambda)^2}
                              {1 + (\tau_\lambda \omega_\lambda)^2}
       \bigg(\mbf{v}_B + \frac{1}{\tau_\lambda \omega_\lambda} \,
                           \frac{\mbf{B}}{B} \mbf{\times v}_B \bigg).
                                  \label{vlambdavec1}
\end{equation}
Elimination of $\mbf{v}_B$ by using equation (\ref{vbvec}) gives
\begin{equation}
   \mbf{v}_{\lambda\perp} = \frac{(\tau_\lambda \omega_\lambda)^2}
                              {1 + (\tau_\lambda \omega_\lambda)^2}
       \bigg(\frac{c}{B}\bigg)^2 \frac{1}{\sigma_{\rm P}^2 + \sigma_{\rm H}^2}
   \bigg[ \bigg(\sigma_{\rm P}
               + \frac{\sigma_{\rm H}}{\tau_\lambda \omega_\lambda} \bigg)
            \frac{1}{c}\, \mbf{j \times B}
        + \bigg(\frac{\sigma_{\rm P}}{\tau_\lambda \omega_\lambda}
                 -\sigma_{\rm H} \bigg) \frac{\mbf{B}}{B} \mbf{\times}
            \frac{1}{c}\, (\mbf{j \times B}) \bigg].  \label{vlambdavec2}
\end{equation}
It is easy to confirm that equation (\ref{vlambdavec2}) is identical with 
equations (\ref{vlx}) and (\ref{vly}).

With some manipulation equation (\ref{current}) can be solved for 
$\mbf{E}_0$ as
\begin{equation}
   \mbf{E}_0 = - \frac{1}{c}\, \mbf{u_{\rm n} \times B}
               + \frac{1}{\sigma_\parallel} \, \mbf{j}
       + \frac{\sigma_{\rm H}}{\sigma_{\rm P}^2 + \sigma_{\rm H}^2} \,
                          \mbf{j \times} \frac{\mbf{B}}{B}
       + \bigg(\frac{\sigma_{\rm P}}{\sigma_{\rm P}^2 + \sigma_{\rm H}^2}
                 - \frac{1}{\sigma_\parallel} \bigg) \frac{\mbf{B}}{B}
              \mbf{\times} \bigg(\mbf{j \times} \frac{\mbf{B}}{B} \bigg).
                              \label{e0a}
\end{equation}
By comparing the coefficients we can confirm that equations (\ref{e0}) 
and (\ref{e0a}) are identical.  Desch \& Mouschovias (2001) used an 
equation equivalent to ours (\ref{e0}) or (\ref{e0a}) though neglecting 
the dominant terms of $\sigma_{\rm P}$ and $\sigma_{\rm H}$ as pointed out 
in \S\,\ref{contgrain}.

\section{Effect of Collision between Charged Particles} \label{collision}

In \S\,\ref{drift} and Appendix~\ref{anotherform} we have neglected 
frictional forces on charged particles except those exerted by the 
neutrals.  Except for electrons in some limited situations, this is a 
good approximation.  We discuss the effect of collision between electrons 
and other charged particles on the collision time $\tau_{\rm e}$ of 
electrons, which appears in the electric conductivity $\sigma_{\rm e}$ 
in equation (\ref{sigmapara}).

The $90^\circ$ deflection time of an electron of velocity $v_{\rm e}$ 
by collision with ions of electric charge $q_{\rm i} = e$ is given by
\begin{equation}
   \tau_{\mbox{\scriptsize e-i}} = \frac{m_{\rm e}^2 v_{\rm e}^3}
                   {8\pi n_{\rm i}\,e^4 \ln {\mit\Lambda}}, \label{tauei}
\end{equation}
where $\ln \mit\Lambda$ is a quantity determined by the cut-off of the 
impact parameter for the collision (\cite{spi62}).  For $v_{\rm e}$ 
we take the thermal velocity of electrons.  The collision time of an 
electron with the neutrals is given by
\begin{equation}
   \tau_{\mbox{\scriptsize e-n}} \approx \frac{1}
      {[\,n({\rm H}_2)\,\sigma(\mbox{e\,-\,H}_2) + 
        n({\rm He}) \,\sigma(\mbox{e\,-\,He})\,]\,v_{\rm e}}, \label{tauen}
\end{equation}
where $\sigma(\mbox{e\,-\,H}_2) \approx 6.6 \times 10^{-16}\,{\rm cm}^2$ 
and $\sigma(\mbox{e\,-\,He}) \approx 2.9 \times 10^{-16}\,{\rm cm}^2$ are 
the collision cross sections of an electron with an H$_2$ molecule and 
an He atom, respectively, at $T \approx 10$\,K (\cite{san00}).  
The collision time $\tau_{\rm e}$ of an electron with ions or neutrals is 
given by
\begin{equation}
   \frac{1}{\tau_{\rm e}} = \frac{1}{\tau_{\mbox{\scriptsize e-i}}}
                            + \frac{1}{\tau_{\mbox{\scriptsize e-n}}}.  
                                          \label{taue}
\end{equation}

From equations (\ref{tauei}) and (\ref{tauen}) we have
\begin{equation}
   \frac{\tau_{\mbox{\scriptsize e-i}}}{\tau_{\mbox{\scriptsize e-n}}}
   \approx \bigg(\frac{n_{\rm i}/n_{\rm H}}{1 \times 10^{-10}} \bigg)^{-1}
           \bigg(\frac{\ln {\mit\Lambda}}{40} \bigg)^{-1} \label{taueitauen}
\end{equation}
for $T \approx 10$\,K.  Thus, the electric conductivity $\sigma_\parallel$ 
is significantly affected by collision of electrons with ions 
only when $n_{\rm i} / n_{\rm H} \gtrsim 1 \times 10^{-10}$, or at 
$n_{\rm H} \lesssim 1 \times 10^8 \,{\rm cm}^{-3}$ as seen from Figure 1.  
Although charged grains also scatter electrons, their contribution does not 
affect this conclusion much because 
$n({\rm g}^+) \ll n({\rm g}^-) \lesssim n_{\rm i}$ at 
$n_{\rm H} \lesssim 1 \times 10^8 \,{\rm cm}^{-3}$, 
$[n({\rm g}^-) + n({\rm g}^+)]/ n_{\rm H} \lesssim 1 \times 10^{-10}$ at 
$n_{\rm H} \gtrsim 1 \times 10^8 \,{\rm cm}^{-3}$,  and $\ln \mit\Lambda$ 
for electron-grain collision is not much different from that for 
electron-ion collision.

Similarly ions are scattered by charged grains.  This effect is important 
compared with their collision with the neutrals only when 
$[n({\rm g}^-) + n({\rm g}^+)]/ n_{\rm H} \gtrsim 2 \times 10^{-9}$, 
which is not realized as seen from Figure 1.  Collision of ions with 
electrons has negligible effect on the motion of ions unless 
$n_{\rm e}/n_{\rm H} \gtrsim 4 \times 10^{-7}$.  Collision of charged 
grains with ions is also much less effective than their collision with 
the neutrals.

Although $\sigma_\parallel$ is greatly affected by the scattering of 
electrons by ions at $n_{\rm H} \lesssim 10^8 \,{\rm cm}^{-3}$, 
$\sigma_{\rm P}$ and $\sigma_{\rm H}$ are hardly affected as can be 
confirmed in the following way.  Even if the effect of collision with ions 
is taken into account on  $\tau_{\rm e}$  at 
$n_{\rm H} \lesssim 1 \times 10^8 \,{\rm cm}^{-3}$, we find 
    $|\tau_{\rm e} \omega_{\rm e}| \gg \tau_{\rm i} \omega_{\rm i} > 1$ 
as far as $n_{\rm i} / n_{\rm H} \ll 10^{-6}$, which is satisfied at least 
in the density range covered by this paper.  Therefore, contribution of 
electrons to $A_1$ and $A_2$ (or $\sigma_{\rm P}$ and $\sigma_{\rm H}$) 
is still much smaller than that of ions as can be confirmed from equations 
(\ref{a1}) and (\ref{a2}).  Our results are therefore not affected by the 
correction on $\tau_{\rm e}$.  On the other hand, we have used 
$\sigma_\parallel$ with the corrected $\tau_{\rm e}$ in the estimation of 
the Ohmic dissipation time $t_{\rm od} \approx 10^{15}\,{\rm yr}$ at 
$n_{\rm H} \approx 10^5 \,{\rm cm}^{-3}$ in \S\,1;  without this 
correction, $t_{\rm od}$ at this density will be $10^2$ times larger.

\section{On the Term ``Ambipolar Diffusion"} \label{termin}

Ambipolar diffusion is a term originally used in plasma physics.  At length 
scales larger than the Debye shielding length in plasmas, electrical 
neutrality is well established.  Therefore, if there is a gradient in the 
electron density, they are transported with a flux proportional to their 
density gradient, and ions follow them because of electric forces, and 
vice versa.  Thus, particles of both electric charges diffuse as one unit.  
This process is called ambipolar diffusion (e.g., \cite{cap76}), whose 
literal meaning fits the process.

In the drift of magnetic fields in molecular clouds investigated first by 
Mestel \& Spitzer (1956), ions and electrons, which are well frozen to 
magnetic fields, drift with the same velocity as if they were one unit.  
Because of this seeming similarity this process got the name 
``ambipolar diffusion".  However, the drift is driven by the magnetic 
force, not by the density gradient.  Thus, this process is quite different 
from what the physical term "diffusion" means.

In the original ambipolar diffusion the diffusion coefficient perpendicular 
to magnetic field lines is inversely proportional to both $B^2$ and the 
collision time of electrons with neutrals, $\tau_{\rm e}$, when 
$\tau_{\rm i} \omega_{\rm i} \gg 1$ (\cite{cap76}), and thus the diffusion 
time is proportional to $\tau_{\rm e} B^2$.  On the other hand, the drift 
time of magnetic fields given by equation (\ref{tbambi}) in the same 
situation $\tau_{\rm i} \omega_{\rm i} \gg 1$, 
$t_B \propto \tau_{\rm i}^{-1} B^{-2}$, has the opposite dependences.  
This difference also shows that these processes are quite different from 
each other.

Because of these differences the latter process was sometimes called 
``plasma drift" instead of ambipolar diffusion (e.g., \cite{spi78}; 
\cite{nak84}; \cite{mes99}).

In this paper we have found that grains, the main contributors to the 
frictional force, of opposite charges drift in opposite directions except 
at relatively low densities.  This is even contrary to the literal meaning 
of ambipolar diffusion.  Some appropriate term is desired;  plasma drift 
is much better than ambipolar diffusion.

%\clearpage

%%%UCP%%%
\newpage
\plotone{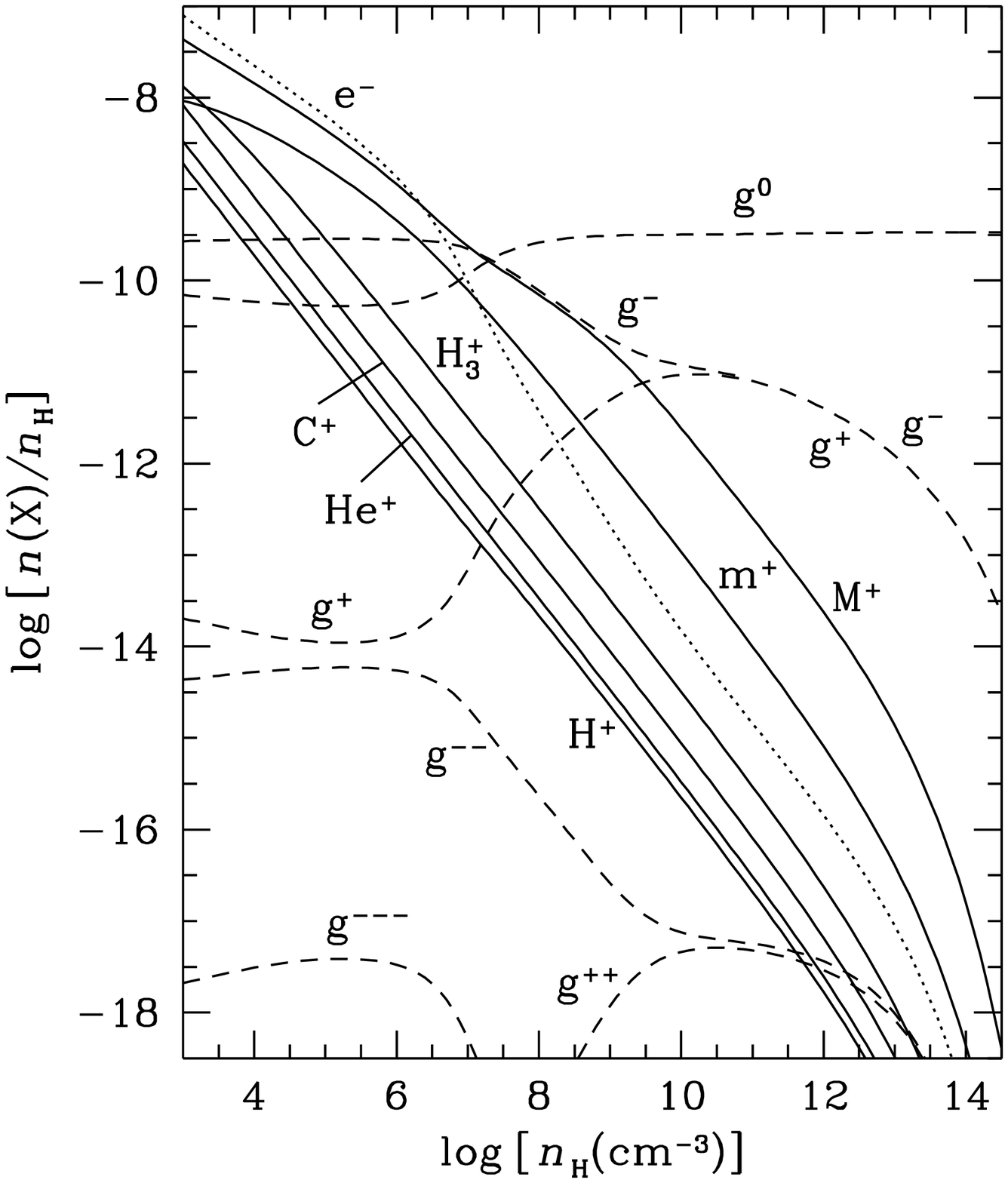}

\figcaption{Abundances of various particles, $n({\rm X})/n_{\rm H}$, as 
functions of the density $n_{\rm H}$ of the cloud by number of hydrogen 
nuclei.  The solid lines are for ions, and the dotted line is for 
electrons.  The dashed lines labeled g$^x$ represent number densities 
relative to $n_{\rm H}$ of grains of charge $xe$ summed up over the radius.  
We have taken the ionization rate of an H$_2$ molecule by cosmic rays 
outside the cloud, $\zeta_0 = 1 \times 10^{-17}\,{\rm s}^{-1}$ (standard 
case).  We have assumed that 20\% of C and O and 2\% of metallic elements 
remain in the gas phase and the rest in grains. \label{fig1}}

%\plottwo{f1.eps}{f2.eps}

\newpage
\plotone{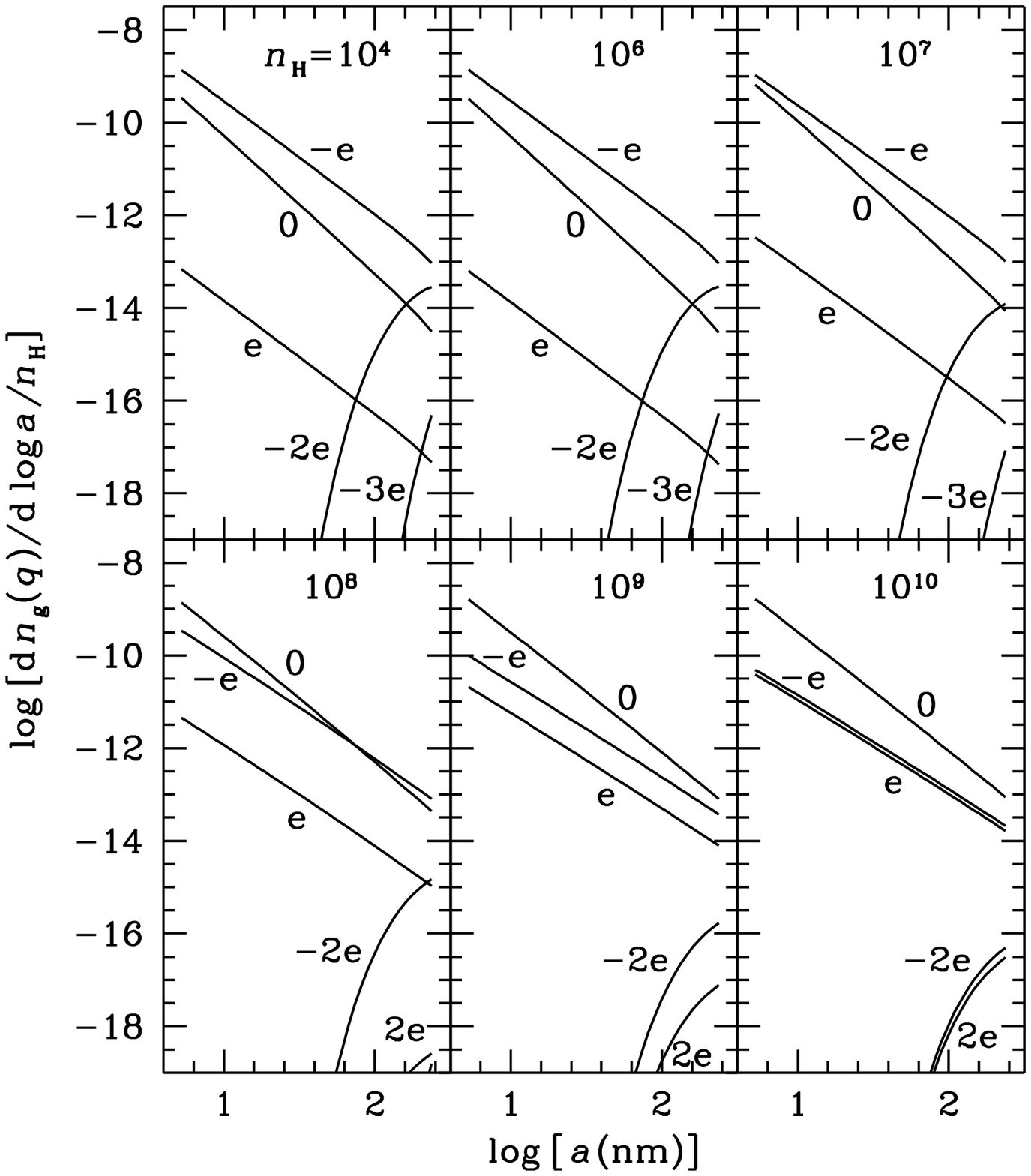}

\figcaption{The charge state distribution of grains as a function of the 
grain radius $a$ at several cloud densities for the same parameters as in 
Fig. 1.  Each panel is labeled with $n_{\rm H}$ in ${\rm cm}^{-3}$, and 
each line is labeled with the grain charge $q_{\rm g}$. \label{fig2}}

\newpage
\plotone{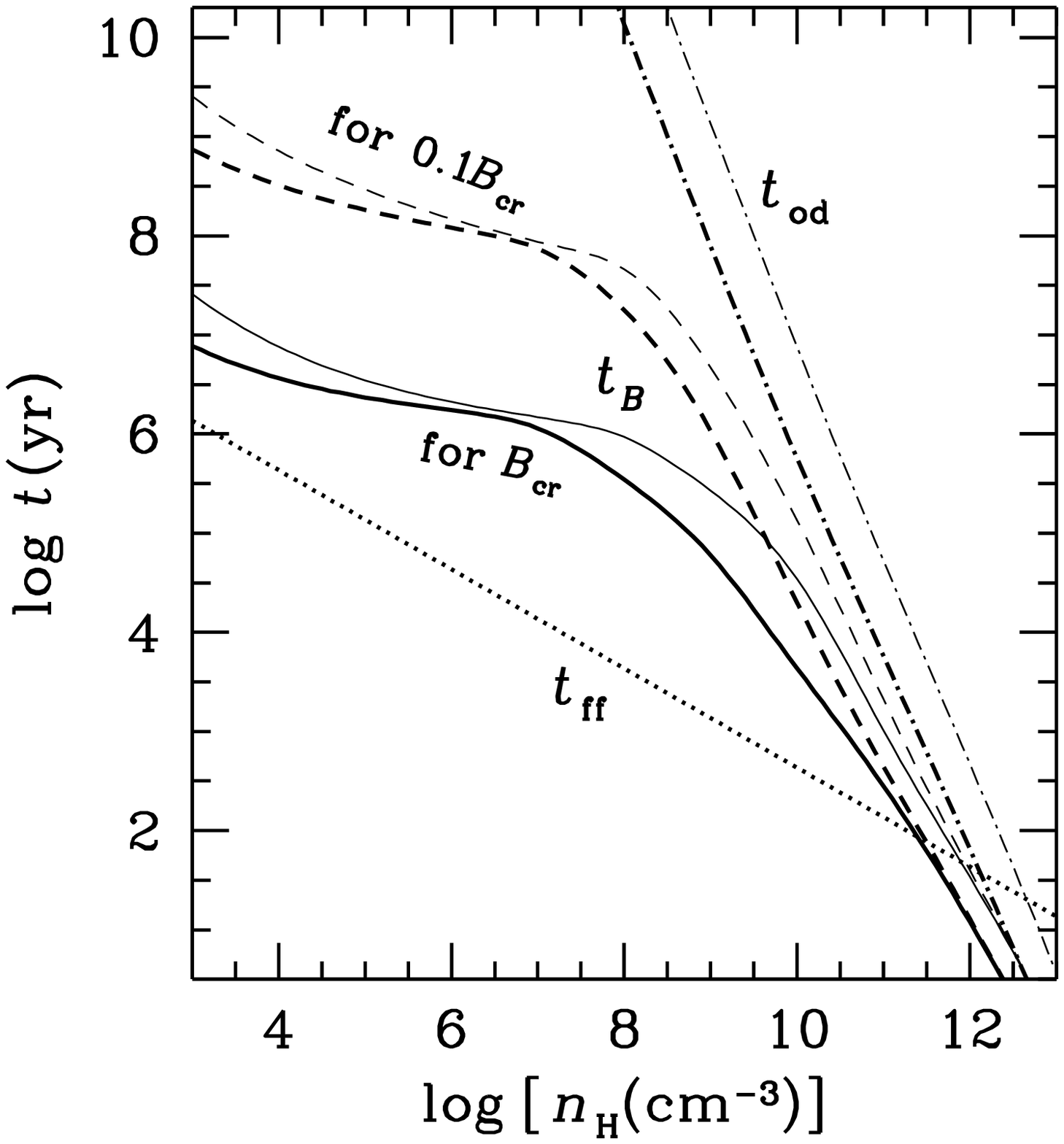}

\figcaption{Time scales of magnetic flux loss for the clouds in which 
force balance approximately holds along field lines, or $B_{\rm cr}$ is 
given by equation (\protect\ref{bcra}).  The flux loss time $t_B$ is shown 
for the two cases of field strength $B = B_{\rm cr}$ ({\it solid lines}) 
and $B = 0.1 B_{\rm cr}$ ({\it dashed lines}).  The Ohmic dissipation time 
$t_{\rm od}$ is shown by the dot-dashed lines.  Two cases of the 
ionization rate by cosmic rays, $\zeta_0 = 1 \times 10^{-17}\,{\rm s}^{-1}$ 
({\it thick lines}: standard case) and $1 \times 10^{-16}\,{\rm s}^{-1}$ 
({\it thin lines}), are shown.  The other parameters are the same as in 
Fig. 1.  For comparison the free-fall time 
$t_{\rm ff} = (3\pi/32G\rho)^{1/2}$ is shown by the dotted line. \label{fig3}}

%\plottwo{f3.eps}{f4.eps}

\newpage
\plotone{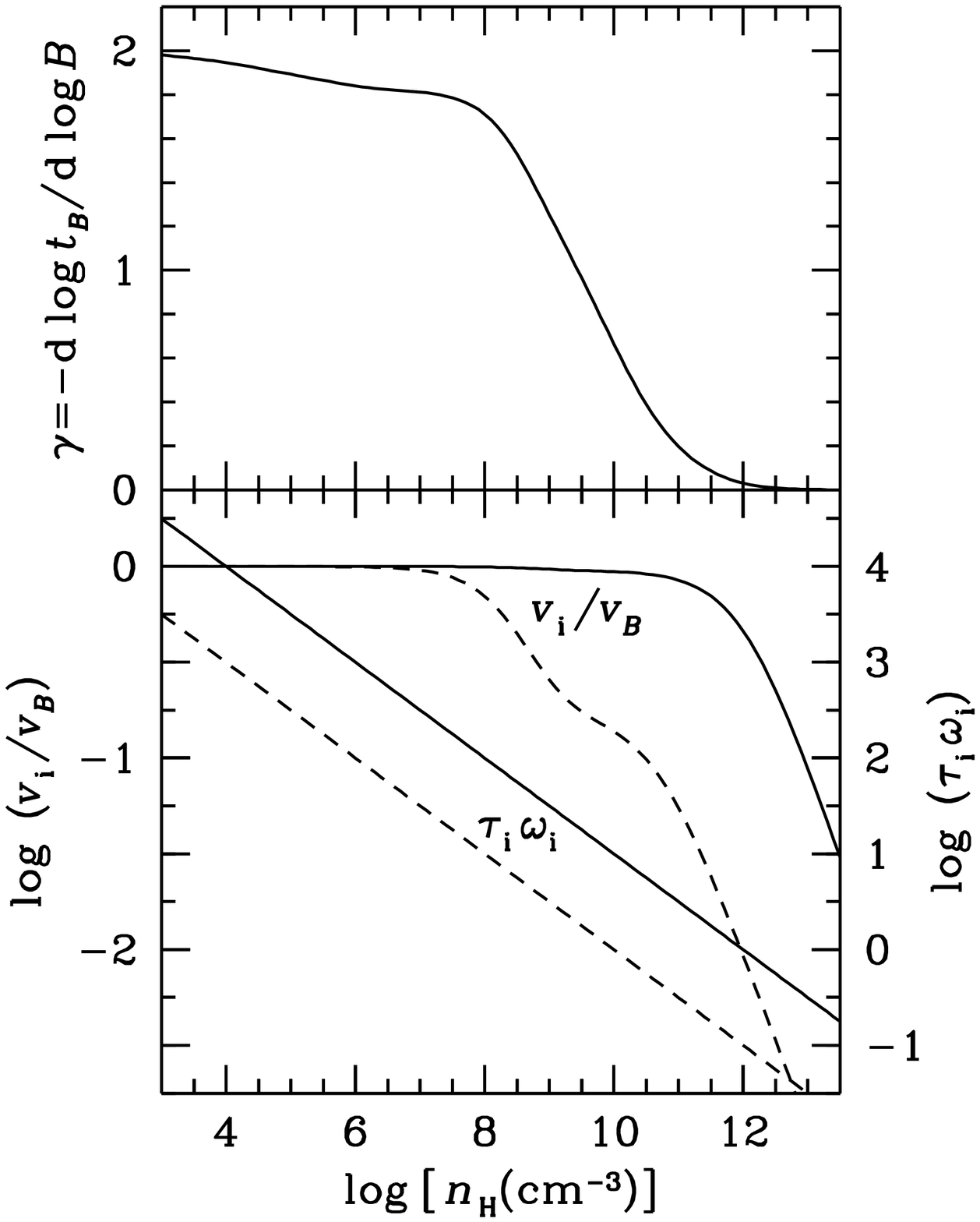}

\figcaption{Dependence of the magnetic flux loss time $t_B$ on the field 
strength $B$ and the drift velocity of ions for the standard case 
$\zeta_0 = 1 \times 10^{-17}\,{\rm s}^{-1}$.  We approximate $t_B$ by a 
power law $t_B \propto B^{-\gamma}$.  
{\it Top:}  The power index $\gamma$ obtained by comparing $t_B$ for the 
two cases $B /B_{\rm cr} = 1$ and 0.1 shown in Fig. 3, 
$\gamma = -{\mit\Delta} \log t_B / {\mit\Delta} \log B$, 
as a function of the cloud density.  For weaker magnetic fields we would 
obtain smaller values of $\gamma$.  
{\it Bottom:}  The drift velocity in the direction of magnetic force 
relative to that of magnetic fields, $v_{\rm i}/v_B$, given by equation 
(\protect\ref{vlvb}), and $\tau_{\rm i} \omega_{\rm i}$ of metallic ions 
M$^+$, dominant among various ions, for the two cases of field strength 
$B = B_{\rm cr}$ ({\it solid line}) and $B = 0.1 B_{\rm cr}$ 
({\it dashed line}).  We have omitted the subscript $x$ to the velocities.  
These quantities take almost the same values for molecular ions m$^+$ 
other than H$_3^+$, abundant next to M$^+$, because their mean mass is not 
much different from that of M$^+$. \label{fig4}}

\newpage
\plotone{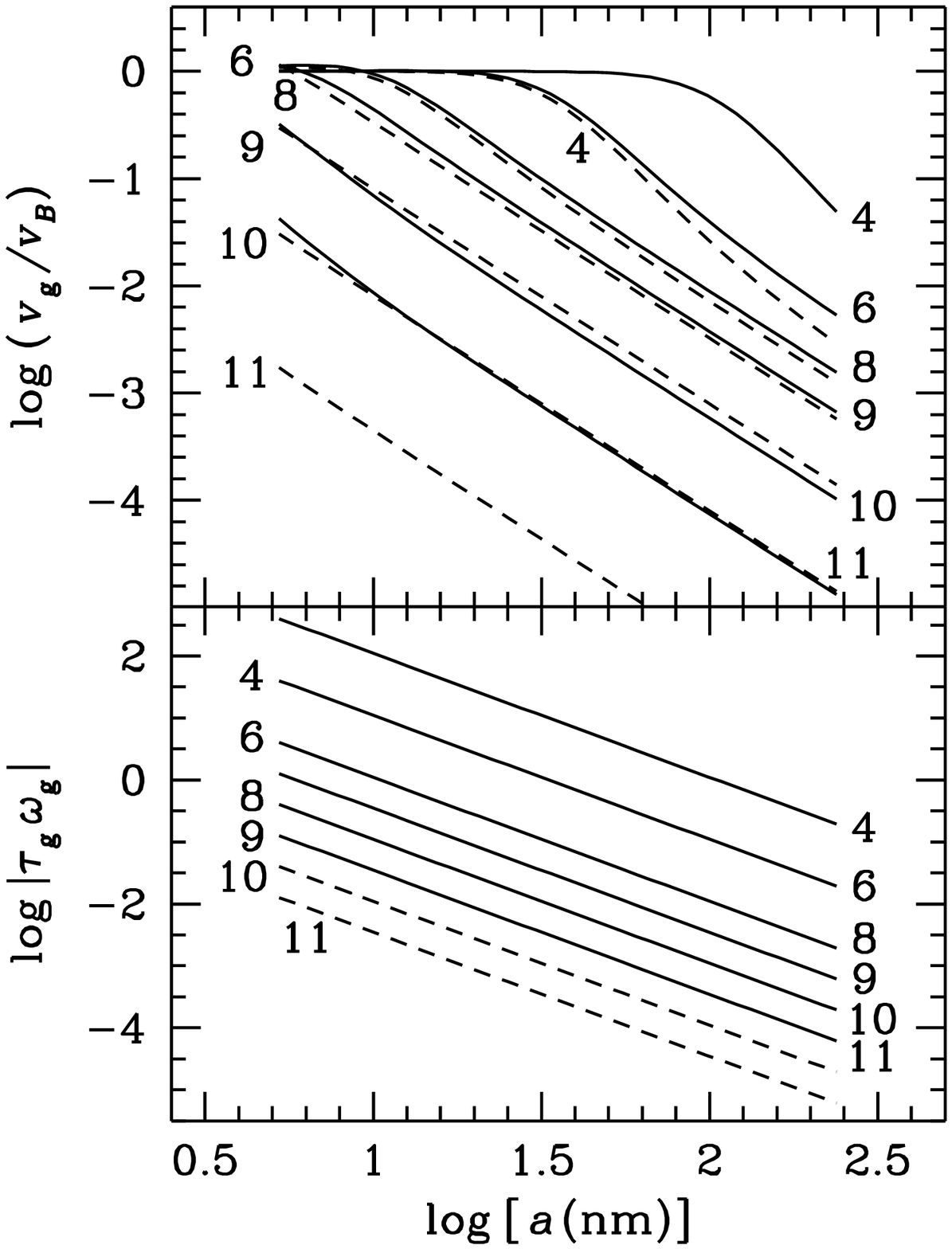}

\figcaption{The drift velocity relative to that of magnetic fields, 
$v_{\rm g}/v_B$ ({\it top}), given by equation (\protect\ref{vlvb}), 
and $|\tau_{\rm g}\omega_{\rm g}|$ ({\it bottom}) for grains of charge 
$-e$ as functions of the grain radius $a$ at several densities for the 
same case as in Fig. 4.  The two cases of field strength $B = B_{\rm cr}$ 
({\it solid lines}) and $B = 0.1 B_{\rm cr}$ ({\it dashed lines}) are 
shown though the dashed lines for $\tau_{\rm g}\omega_{\rm g}$ overlap 
with the solid lines except for $\log n_{\rm H} = 10$ and $11$ in 
${\rm cm}^{-3}$.  The values of $\log n_{\rm H}$ are attached to the 
right ends of the lines for $B = B_{\rm cr}$ and mostly to the left ends 
for $B = 0.1 B_{\rm cr}$. \label{fig5}}

%\plottwo{f5.eps}{f6.eps}

\newpage
\plotone{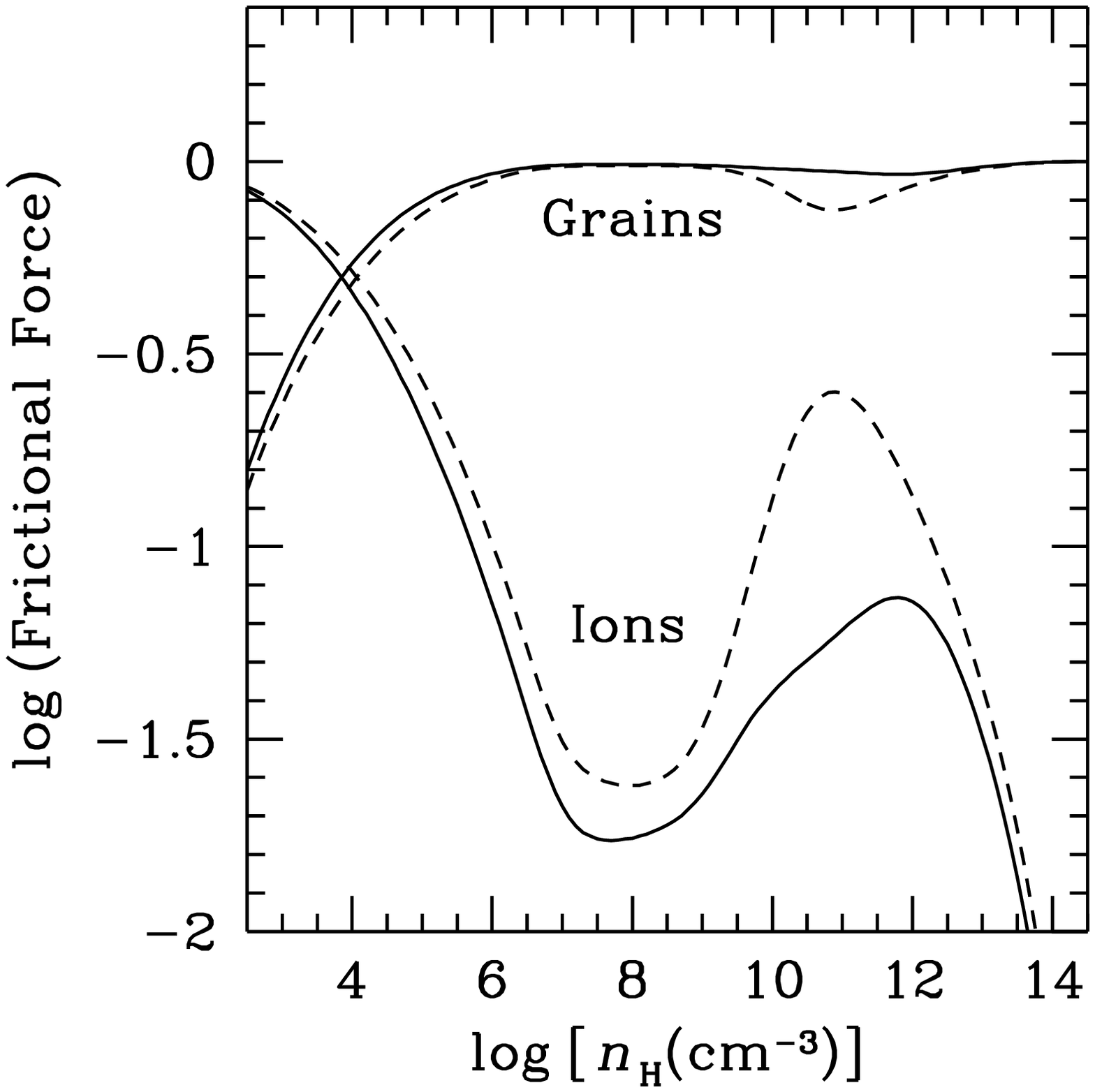}

\figcaption{Frictional forces exerted by ions and by grains on the neutrals 
relative to the total frictional force, which is equal to 
$|\,\protect\mbf{j \times B}|/c$ per unit volume, as functions of the cloud 
density $n_{\rm H}$ for the same case as in Fig. 4.  Each kind of particles 
exerts the frictional force given by each term of equation 
(\protect\ref{frictionx}).  The frictional forces were summed up for all 
kinds of ions and for grains of all radii and all charges.  Shown are the 
two cases of the field strength $B = B_{\rm cr}$ ({\it solid lines}) and 
$B = 0.1B_{\rm cr}$ ({\it dashed lines}).  The frictional force exerted by 
electrons is more than 3 orders of magnitude smaller than that by ions. 
\label{fig6}}

\newpage
\plotone{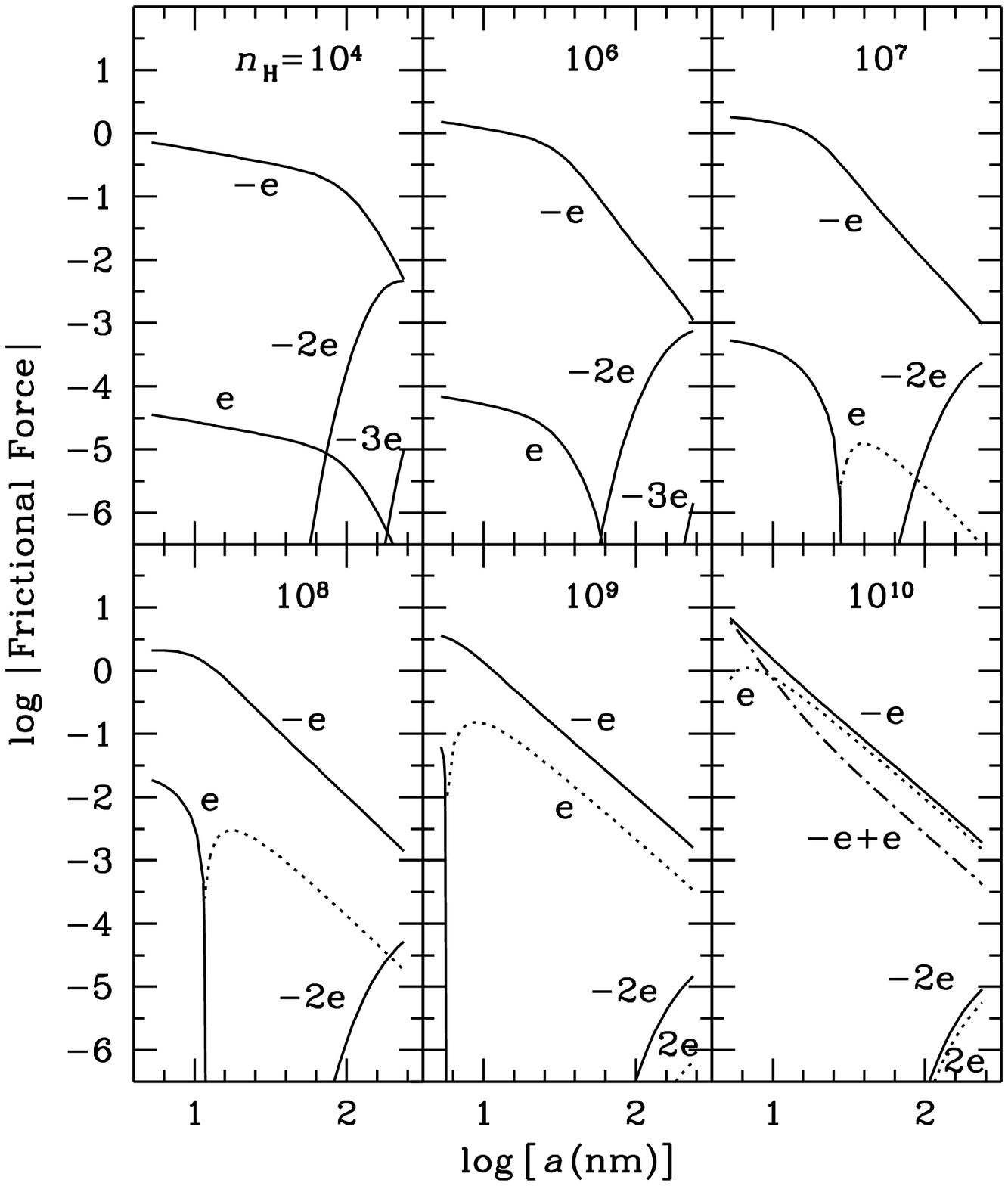}

\figcaption{The frictional force exerted by grains on the neutrals as a 
function of their radius $a$ and electric charge $q_{\rm g}$ at several 
cloud densities for the same case as in Fig. 4 with $B = B_{\rm cr}$.  
Each curve shows the frictional force per unit logarithmic radius width 
${\mit\Delta} \log a = 1$ relative to the total frictional force, 
$|\,\protect\mbf{j \times B}|/c$ per unit volume.  Each panel is labeled 
with the density $n_{\rm H}$ in cm$^{-3}$, and each line is labeled with 
the grain charge $q_{\rm g}$.  The {\it solid lines} are when the frictional 
force is parallel to the magnetic force, and the {\it dotted lines} are when 
it is anti-parallel to the magnetic force, apart from the component 
perpendicular to \,$\protect\mbf{j \times B}$.  The {\it dot-dashed line} 
in the panel of $n_{\rm H} = 10^{10}\,{\rm cm}^{-3}$ shows the sum of the 
frictional forces exerted by grains of charge $-e$ and $e$. \label{fig7}}

%\plottwo{f7.eps}{f8.eps}

\newpage
\plotone{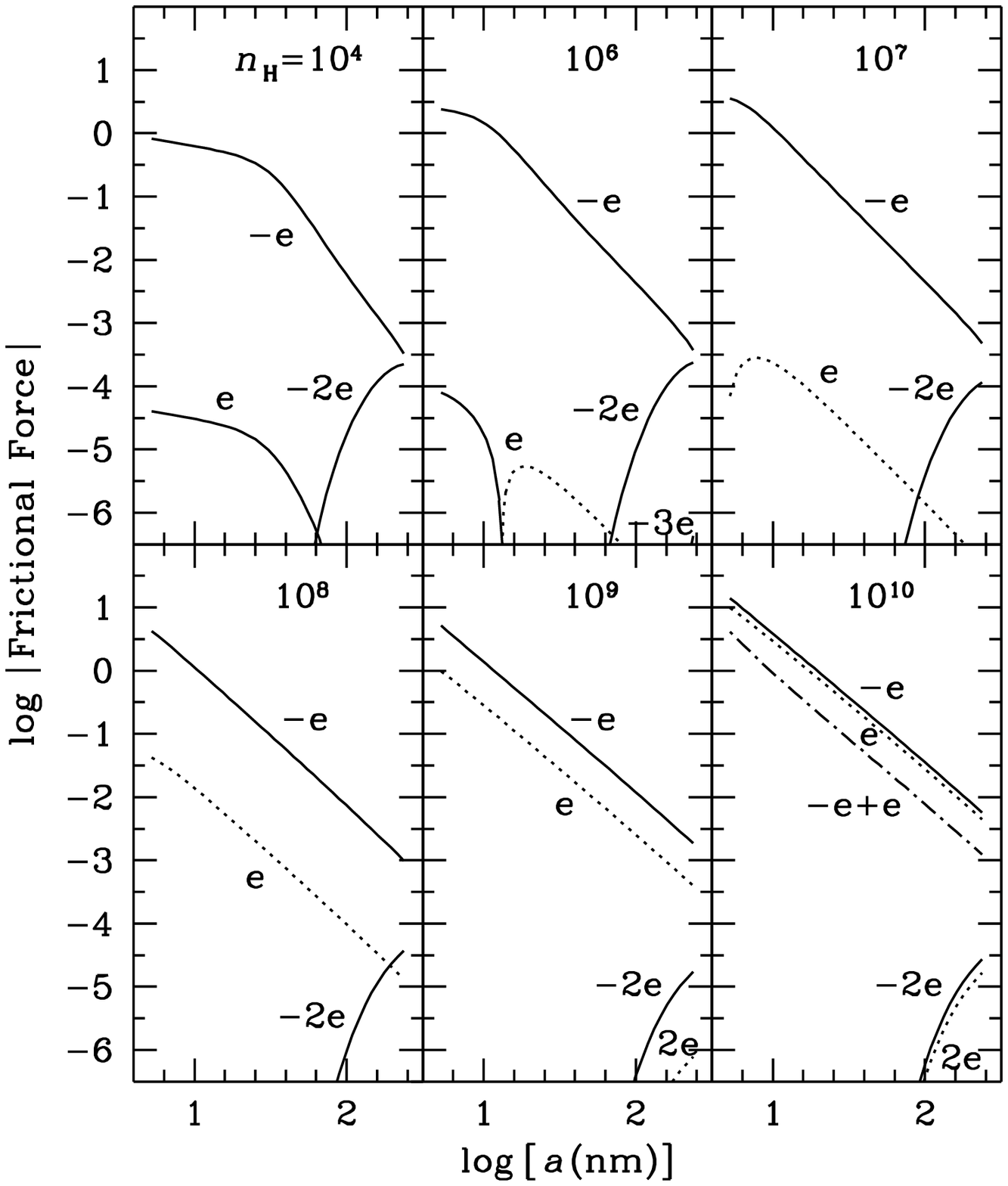}

\figcaption{Same as Fig. 7 but for $B = 0.1 B_{\rm cr}$. \label{fig8}}

\newpage
\plotone{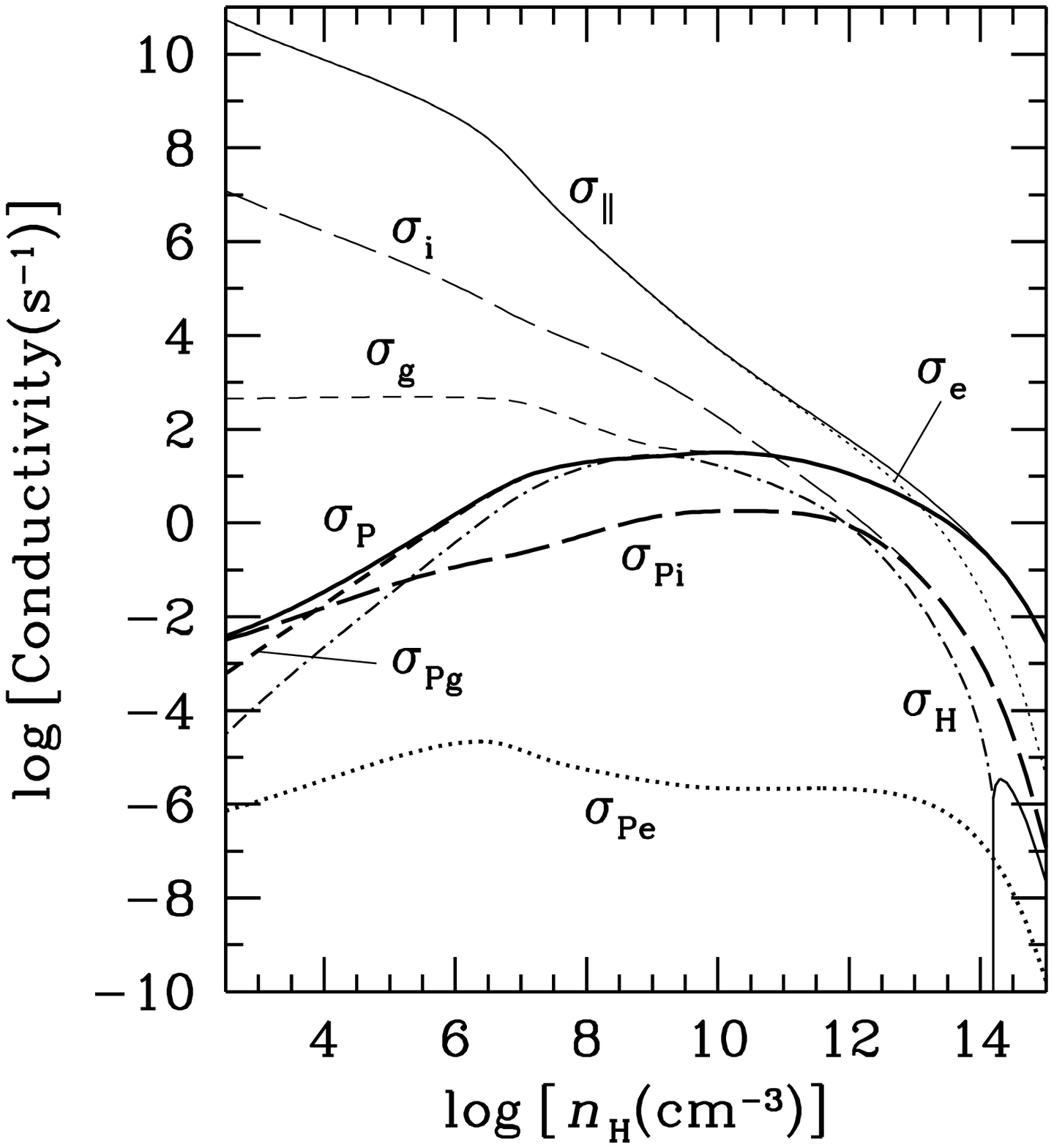}

\figcaption{Electric conductivities as functions of the mean density of 
the cloud, $n_{\rm H}$, for the same case as in Fig. 4.  The thin lines 
are for conductivity $\sigma_\parallel$ along magnetic field lines, 
and the thick lines are for Pedersen conductivity $\sigma_{\rm P}$ for 
the case of $B = B_{\rm cr}$.  Contributions of electrons, ions, and 
grains are shown by dotted, long-dashed, and short-dashed lines, 
respectively, and the totals are shown by solid lines.  The line labeled 
$\sigma_{\rm H}$ represents absolute values of Hall conductivity for 
the case of $B = B_{\rm cr}$; the dot-dashed line is where 
$\sigma_{\rm H} < 0$, and the solid line with $\sigma_{\rm H} > 0$. 
\label{fig9}}

\end{document}